\documentclass[11pt,onecolumn,draft]{IEEEtran} 
\usepackage{cite}
\usepackage{psfig}
\usepackage{subfigure}
\usepackage{graphicx}
\usepackage{amsmath}
\usepackage{amssymb}

\newcommand{\bb}[1]{{\mathbf{#1}}}

\newcommand{\nn}{\nonumber}

 \DeclareMathOperator{\diag}{diag}
\DeclareMathOperator{\tr}{tr}

\newtheorem{theorem}{Theorem}

\newtheorem{definition}{Definition}

\begin{document}

\title{Spectrum Sensing in Wideband OFDM Cognitive Radios}

\author{
\begin{tabular}{ccc}
  Chien-Hwa Hwang & & Shih-Chang Chen \\
  \texttt{chhwang@ee.nthu.edu.tw} & & \texttt{iamcsc@realtek.com.tw}\\
  Inst. of Commun. Engr., & &  Digital IC Design Dept.\\
  National Tsing Hua University, & &  Realtek Semiconductor Corp.,\\
  Hsinchu, Taiwan. & &  Hsinchu, Taiwan.\\
\end{tabular}
}

\footnotetext[1] {Chien-Hwa Hwang is the author for correspondence.}

\maketitle

\begin{abstract}

In this paper, detection of the primary user (PU) signal in an
orthogonal frequency division multiplexing (OFDM) based cognitive
radio (CR) system is addressed. According to the prior knowledge of
the PU signal known to the detector, three detection algorithms
based on the Neyman-Pearson philosophy are proposed. In the first
case, a Gaussian PU signal with completely known probability density
function (PDF) except for its received power is considered. The
frequency band that the PU signal resides is also assumed known.
Detection is performed individually at each OFDM sub-carrier
possibly interfered by the PU signal, and the results are then
combined to form a final decision. In the second case, the
sub-carriers that the PU signal resides are known. Observations from
all possibly interfered sub-carriers are considered jointly to
exploit the fact that the presence of a PU signal interferers all of
them simultaneously. In the last case, it is assumed no PU signal
prior knowledge is available. The detection is involved with a
search of the interfered band. The proposed detector is able to
detect an abrupt power change when tracing along the frequency axis.

\end{abstract}

\newpage

\section{Introduction}

Radio spectrum is the medium for all types of wireless
communications, such as cellular phones, satellite-based services,
wireless low-powered consumer devices, and so on. Since most of the
usable spectrum has been allocated to existing services, the radio
spectrum has become a precious and scarce resource, and there is an
urgent concern about the availability of spectrum for future needs.
Nonetheless, the allocated radio spectrum today is not efficiently
utilized. According to a report of the United States Federal
Communications Commission (FCC) \cite{FCC1}, there are large
temporal and geographic variations in the utilization of allocated
spectrum ranging from 15\% to 85\%. Moreover, according to Defense
Advanced Research Projects Agency (DARPA), in the United States,
only 2\% of the spectrum is in use at any moment. It is then clear
that the solution to the spectrum scarcity problem is dynamically
looking for the spectrum "white spaces" and using them
opportunistically. Cognitive radio (CR) technology, defined first by
J. Mitola \cite{mitola99,mitola00}, is thus advocated by FCC as a
candidate for implementing opportunistic spectrum sharing. The
spectrum management rule of CR is that all new users for the
spectrum are secondary (cognitive) users and requiring that they
must detect and avoid the primary user.

To achieve the goal of CR, it is a fundamental requirement that the
cognitive user (CU) performs spectrum sensing to detect the presence
of the primary user (PU) signal. Digital signal processing
techniques can be employed to promote the sensitivity of the PU
signal sensing. Three commonly adopted methods are matched
filtering, energy detection
\cite{urkowitz67,kostylev02,digham03,green05,ghasemi05,cabric06,mishra06},
and PU signal feature detection with the cyclo-stationary feature
most widely adopted \cite{oner07,ghozzi06,han06,lunden07}. Moreover,
cooperation among CUs in spectrum sensing can not only reduce the
detection time and thus increase the agility, but also alleviate the
problem that a CU fails to detect the PU signal because it is
located at a weak-signal region
\cite{ghasemi05,cabric06,mishra06,visotsky05,weiss03,ganesan05,ganesan07,ganesan07_1,gandetto07}.
For overview of these approaches and their properties, see
\cite{cabric04,mishra2007,akyildiz06}.

It is concluded in \cite{tang05} that orthogonal frequency division
multiplexing (OFDM) is the best physical layer candidate for a CR
system since it allows easy generation of spectral signal waveforms
that can fit into discontinuous and arbitrary-sized spectrum
segments. Besides, OFDM is optimal from the viewpoint of capacity as
it allows achieving the Shannon channel capacity in a fragmented
spectrum. Owing to these reasons, in this paper, we conduct spectrum
sensing in an OFDM based CR system.

In detection theory, the Neyman-Pearson (NP) criterion is used when
there is difficulty in determining the prior probabilities and
assigning costs \cite{trees01} for hypotheses, which is the case in
our PU signal detection. The NP detector compares the likelihood
ratio (LR) with a threshold determined by the constraint of false
alarm probability to decide which hypothesis is true. However, in
many cases, some PU signal parameters, such as power level,
correlation properties, frequency band, and so on, may not be known.
At this moment, the PU signal detection problem becomes a composite
hypothesis testing, which requires performing estimation for those
unknown parameters in the probability density function (PDF) of the
observation for either hypothesis. Thus, the degree of detector
complexity is directly related to the knowledge of the signal and
noise characteristics in terms of their PDFs. Moreover, since the
estimation error is not negligible, the detection performance
decreases as we have less specific knowledge of the signal and noise
characteristics. According to the prior knowledge about the PU
signal, three cases of PU signal detection in a cognitive OFDM
system are considered in this paper.

In Case A, we assume the PU signal model is known and consider a
Gaussian PU signal with completely known PDF except for its received
power. The normalized, i.e. unity diagonal elements, covariance
matrix of the PU signal can be derived directly from the model
assumed for it. As the received power as well as the normalized
covariance matrix of the PU signal are distinct at each OFDM
sub-carrier, PU signal detection in this case is executed
individually at each sub-carrier, and the results are then combined
together to form a final decision. In Case B, neither the model of
the PU signal nor its distribution is known to the detector. The
prior knowledge is the frequency band that the PU signal resides.
The band is assumed to be a continuous segment of sub-carriers. To
incorporate the fact that, once a PU signal occurs, several
sub-carriers in a row are interfered simultaneously, the detector
makes its decision by jointly considering observations from all
possibly interfered sub-carriers. In Case C, no prior knowledge of
the PU signal is available. Thus, the detection is involved with a
search of possibly interfered band. The proposed detector is able to
detect an abrupt power change when tracing along sub-carriers.

The organization of this paper is summarized as follows. In Section
II, the signal model of a cognitive OFDM system interfered by a PU
signal is derived. Three cases concerning the PU signal prior
knowledge are also described. In Section III, the designs of PU
signal detectors are carried out with PU signal prior information
stated in Section II. Simulation results of the proposed detection
algorithms are given in Section IV. Finally, we conclude this paper
in Section V.

\section{PU Signal Detection in a Cognitive OFDM System}

Consider a wideband cognitive OFDM system with $Q$ sub-carriers. The
binary data stream generated from the source is encoded and
interleaved, and then subdivided into groups of $B$ bits used to
generate blocks of $Q$ symbols, where each symbol assumes one of $L$
possible values with $B=Q \log_2 L$. It is assumed that $(\log_2
L)$-ary phase shift keying (PSK) modulation is employed. We denote
the constellation points corresponding to the $n$-th block of $Q$
symbols by ${\cal S}(n)=\{S_0(n),S_1(n),\cdots,S_{Q-1}(n)\}$. The
$n$-th OFDM symbol is generated by feeding ${\cal S}(n)$ into a
$Q$-point inverse discrete Fourier transform (IDFT) and
pre-appending the output with cyclic prefix (CP). The resultant
signal is up-converted to the carrier frequency, and then
transmitted over a wireless fading channel.

At the receiver, after the frequency down conversion and the CP
removal, the output signal is passed through a $Q$-point discrete
Fourier transform (DFT). In the presence of a PU signal, the DFT
output corresponding to the $n$-th OFDM symbol is given by
\begin{equation}\label{eq:mouse0530}
Y_q(n)=H_q(n) \cdot S_q(n)+I_q(n)+W_q(n),\quad 0\leq q\leq Q-1,
\end{equation}
where $H_q(n)$ is the frequency response of the channel at
sub-carrier $q$ experienced by the $n$-th OFDM symbol, and
$\{I_q(n)\}$ and $\{W_q(n)\}$ are the contributions resulting from
the PU signal and additive white Gaussian noise (AWGN),
respectively.

Suppose that a PU signal occupies the frequency band extending from
the $q_0$-th to the $q_1$-th sub-carriers of the OFDM system. If the
information of the PU signal frequency band, i.e. $q_0$ and $q_1$,
is known to the detector, the detection algorithm decides whether
the signal $\{I_q(n)\}$ is present in (\ref{eq:mouse0530}) based on
the observation $\{Y_q(n):0\leq n\leq N-1,q_0\leq q\leq q_1\}$,
where $N$ is the observation length at each sub-carrier, and, if
any, the prior knowledge of the PU signal. When $q_0$ and $q_1$ are
not known, the observation $\{Y_q(n)\}$ needs to be extended to all
sub-carriers $0\leq q\leq Q-1$.

\begin{table}
\caption{Prior knowledge of the PU signal}
\begin{center}
\begin{tabular}{lcccc}
\hline & \textit{Received Power} & \textit{Signal Model} &
\textit{Gaussian Distribution} & \textit{Frequency Band} \\
\hline\hline
\textbf{\textit{Case A}} & No & Yes & Yes & Yes\\
\textbf{\textit{Case B}} & No & No & Not necessarily & Yes\\
\textbf{\textit{Case C}} & No & No & Not necessarily & No\\
\hline
\end{tabular}\label{tab:1}
\end{center}
\end{table}

TABLE~\ref{tab:1} lists three cases regarding the amount of prior
knowledge about the PU signal, including the received power, the
signal model, probability distribution, and the frequency band it
resides. In all three cases, the received power of the PU signal is
unknown. In Case A, it is assumed the model of PU signal is known.
Examples that the PU signal characteristic is known to the detector
can be found in, e.g.
[\citenum{mishra2007},\citenum{srikanteswara07},\citenum{oner07}].
We assume the sub-carrier indices $[q_0,q_1]$ occupied by the PU
signal are known, the stochastic process $\{I_q(n)\}$ observed at
each sub-carrier $q_0\leq q\leq q_1$ is Gaussian, and the $N\times
N$ normalized covariance matrices $\bb C_q$'s of the random signal
$\{I_q(n)\}_{n=0}^{N-1}$ at $q_0\leq q\leq q_1$ can be obtained from
the PU signal model. The normalization factor to obtain $\bb C_q$ is
the PU signal received power at that sub-carrier, and $\bb C_q$ has
diagonal components equal to one. In Case B, the assumptions of
known PU signal model and Gaussian distribution are removed. It will
be clear this case serves as an intermediate stage for developing
the detector in Case C, where no prior PU signal knowledge is
available.

\section{Design of PU Signal Detector}

\subsection{Case A: Known PU Signal Model, Probability Distribution, and Frequency
Band}\label{section:3_A}

Two PU signal models, i.e. a sum of tonal signals and an
auto-regressive (AR) stochastic process, are used as examples for
the detection problem. The PU signal $\{I_q(n)\}_{n=0}^{N-1}$ seen
at the $q$-th OFDM sub-carrier has a covariance matrix $P_I(q) \bb
C_q$, where $P_I(q)$ is the unknown received power of the PU signal
at sub-carrier $q$, and $\bb C_q$ is the normalized covariance
matrix with unit diagonal elements.

\subsubsection{Tonal PU Signal}\label{subsection:2.1}

Here we model the PU signal as the sum of a number of complex
sinusoids. Examples include the worldwide interoperability for
microwave access (WiMAX) and wireless local area network (WLAN)
systems, which also employ OFDM technologies. With this model, the
received PU signal is $i(t)=\sum_{l=-\infty}^\infty
i_l(t-lT_i)$\footnote{Here $i(t)$ is the time-domain PU signal,
whereas $\{I_q(n)\}$ given in (\ref{eq:mouse0530}) is a
frequency-domain signal.}, where $T_i$ is the symbol duration, and
$i_l(t)$ is the signal containing the $l$-th symbol. We have
\begin{equation}\label{eq:cow0614}
i_l(t)=\sum_{k=0}^{K-1}\Re\{d_{l,k}(t)e^{j (2\pi
f_{i}t+\phi)}\},\quad 0\leq t\leq T_i,
\end{equation}
where $K$, $f_{i}$ and $\phi$ are the number of complex sinusoids,
the carrier frequency, and the random carrier phase, respectively,
$\Re\{\cdot\}$ denotes the real part, and $d_{l,k}(t)$ is the
complex baseband signal of the $k$-th sinusoid. We have
$$
d_{l,k}(t)=\zeta_{l,k}\cdot X_{l,k} e^{j 2\pi k t/T_i}, \quad 0\leq
t\leq T_i,
$$
where $X_{l,k}$ is the PSK modulated data of the $k$-th sinusoid at
the $l$-th symbol, and $\zeta_{l,k}$ is the channel fading
coefficient of the $k$-th sub-carrier when symbol $l$ of the PU
signal is received. We assume, for each particular $l$, random
variables $\zeta_{l,k}$, $k=0,1,\cdots,K-1$, are identically
distributed.

Let $\eta= \lfloor T_i/T_s\rfloor$, where $T_s$ is the symbol
duration of the cognitive OFDM, $\lfloor x\rfloor$ is the largest
integer no greater than $x$, and
$\beta_{k,q}=[(f_i-f_s+k/T_i)T_s-q]/Q$. It is shown in
Appendix~\ref{appendix:1} that, the $(n,m)$-th element of the
normalized covariance matrix $\bb C_{q}(n,m)$ of
$\{I_q(n)\}_{n=0}^{N-1}$ is given by
\begin{equation}\label{eq:mouse0605}
\bb C_{q}(n,m)=\left\{
\begin{array}{ll}
\left(1-\dfrac{|n-m|}{\eta}\right) \dfrac{{\displaystyle
\sum_{k=0}^{K-1}} e^{j
2\pi(n-m)T_s(f_i+k/T_i)}\dfrac{\sin^2(\pi\beta_{k,q}
Q)}{\sin^2(\pi\beta_{k,q})}}
{{\displaystyle \sum_{k=0}^{K-1}} \dfrac{\sin^2(\pi\beta_{k,q} Q)}{\sin^2(\pi\beta_{k,q})}}, & |n-m|\leq \eta-1,\\
0, & \textrm{otherwise}.
\end{array}
\right.
\end{equation}

\subsubsection{AR PU Signal}\label{subsection:2.2}

We consider the time-domain discrete-time PU signal $\{i_p\}$ at the
output of the sampler following the frequency down-converter.
Suppose that $\{i_p\}$ can be modeled as an $r$-th order AR random
process of
\begin{equation}\label{eq:dragon0908}
i_p=-\sum_{j=1}^r \phi_j i_{p-j}+e_p,
\end{equation}
where $\{e_p\}$ is a white Gaussian random sequence with variance
$\nu^2$, and parameters $\{\phi_j\}_{j=1}^r$ are obtained when
$\{i_p\}$ has unit power. The contribution of the PU signal at the
$q$-th DFT output for $N$ OFDM symbols is $\bb
I_q=[I_q(0),I_q(1),\cdots,I_q(N-1)]^T$, given by
\begin{equation}\label{eq:cow0908}
\bb I_q=\bb F_q\bb i,
\end{equation}
where $\bb F_q=\diag\{\underbrace{\bb f_q,\bb f_q,\cdots,\bb
f_q}_{\textrm{$N$ times}}\}$ with $\bb f_q=[e^{-j2\pi q\cdot
0/Q}\hspace{2mm}e^{-j2\pi q\cdot 1
/Q}\hspace{2mm}\cdots\hspace{2mm}e^{-j2\pi q (Q-1)/Q}]$, and $\bb
i=[i_0, i_{1},$ $\cdots,i_{QN-1}]^T$. It is readily seen that $\bb
I_q$ forms a Gaussian random process as $\{e_p\}$ is modeled to be
Gaussian. In Appendix~\ref{appendix:2}, we show how the normalized
covariance matrix $\bb C_q$ of $\bb I_q$ given in (\ref{eq:cow0908})
can be computed.

It is seen that, for the two PU signal models presented above, the
normalized covariance matrix $\bb C_q$ at each sub-carrier
$q\in[q_0,q_1]$ are distinct. Moreover, the PU signal at various
sub-carriers have different received powers. It will be shown later,
cf. (\ref{eq:snake0908}), these unknown received powers need to be
estimated. Thus, we conduct the PU signal detection individually at
each sub-carrier, and the final decision is made by combining the
individual decisions at sub-carriers. If a joint detection of all
sub-carriers is performed, it is required to estimate all unknown PU
signal powers jointly. To design a detector based on the
Neyman-Pearson philosophy, the detection threshold $\gamma_q$ at
sub-carrier $q$ is determined by a given overall (i.e., combined
from all sub-carriers) false alarm probability
$P_\textrm{FA}=\alpha$ such that the overall detection probability
$P_\textrm{D}$ is maximized. Let the decisions made at individual
sub-carriers be combined by an OR operation, i.e., the detector
decides ${\cal H}_1$ if any of the sub-carriers declares an PU
signal is present. For both $P_\textrm{FA}$ and $P_\textrm{D}$, we
have
\begin{equation}\label{eq:snake0611}
P_\textrm{S}=1-\prod_{q\in [q_0,q_1]}(1-P_\textrm{S}(q)),\quad
\textrm{S}\in\{\textrm{FA,D}\},
\end{equation}
where $P_\textrm{S}(q)$ is the detection or false alarm probability
at sub-carrier $q$. Letting $P_\textrm{FA}(q)$ equal for all $q$'s,
we obtain the false alarm constraint at each sub-carrier as
\begin{equation}\label{eq:tiger0707}
P_\textrm{FA}(q)=1-(1-\alpha)^{1/B_\textrm{PU}},
\end{equation}
where we define the bandwidth of the PU signal as
$B_\textrm{PU}=q_1-q_0+1$.

Suppose that the detection is performed when the cognitive OFDM
system is not transmitting signals. The hypothesis testing at the
$q$-th sub-carrier is
\begin{equation}\label{eq:mouse0611}
\begin{array}{ll}
{\cal H}_0: & Y_q(n)=W_q(n), \\
{\cal H}_1: & Y_q(n)=I_q(n)+W_q(n),
\end{array}\quad n=0,1\cdots,N-1,
\end{equation}
where $\{W_q(n)\}$ is complex white Gaussian noise independent of
$\{I_q(n)\}$ with distribution ${\cal CN}(\bb 0,\sigma_W^2\bb I)$,
and ${\cal H}_0$ and ${\cal H}_1$ represent that the PU signal is
off and on, respectively. The PU signal $\{I_q(n)\}_{n=0}^{N-1}$ is
given in (\ref{eq:mouse0908}) and (\ref{eq:cow0908}), respectively,
when it is modeled as a sum of tonal signals and an AR random
process. Due to the absence of the OFDM signal, $\{Y_q(n)\}$ is a
Gaussian random process in either hypothesis.

Since the detection algorithm proposed for this case is done
individually at sub-carriers, we omit $q$ in $P_I(q)$ for notational
simplicity. The likelihood ratio associated with
(\ref{eq:mouse0611}) is
\begin{equation}\label{eq:horse0908}
L(\bb Y_q)=\dfrac{p(\bb Y_q;P_I,{\cal H}_1)}{p(\bb Y_q;{\cal H}_0)},
\end{equation}
where $\bb Y_q =[Y_q(0),Y_q(1),\cdots,Y_q(N-1)]^T$,
$p(\bb Y_q;P_I,{\cal H}_1)$ is the probability density function
(PDF) of $\bb Y_q$ under ${\cal H}_1$ given as
\begin{equation}\label{eq:mouse0607}
p(\bb Y_q;P_I,{\cal H}_1)=\frac{1}{\pi^{N}\det(P_I\bb
C_q+\sigma_W^2\bb I)}\exp\left(-\bb Y_q^\dag(P_I\bb
C_q+\sigma_W^2\bb I)^{-1}\bb Y_q\right),
\end{equation}
and $p(\bb Y_q;{\cal H}_0)$ is the PDF under ${\cal H}_0$ obtained
by setting $P_I$ in (\ref{eq:mouse0607}) to zero.
Using matrix inversion lemma, we have the test statistic $\ln L(\bb
Y_q)$ expressed by
\begin{equation}\label{eq:horse0908_1}
\ln L(\bb Y_q)=\sigma_W^{-2}P_I \bb Y_q^\dag \bb C_q(P_I\bb
C_q+\sigma_W^2\bb I)^{-1}\bb Y_q-\ln\det\left(P_I\bb
C_q+\sigma_W^{2}\bb I\right)+\ln\sigma_W^{2N}.
\end{equation}
It is seen that the unknown $P_I$ in $(P_I\bb C_q+\sigma_W^2\bb
I)^{-1}$ cannot be decoupled from the observation $\bb Y_q$. Thus,
uniformly most powerful (UMP) test does not exist. Consequently, a
generalized likelihood ratio test (GLRT) is employed, where $P_I$ in
(\ref{eq:horse0908_1}) is replaced with its maximum likelihood (ML)
estimate $\hat{P}_I$.

Let $\bb C_q$ be eigen-decomposed as $\bb C_q=\bb V_q\bb\Lambda_q
\bb V_q^\dag$, where $\bb V_q=[\bb v_{q,0}\hspace{1mm}\bb
v_{q,1}\hspace{1mm}\cdots\hspace{1mm}\bb v_{q,N-1}]$ and
$\bb\Lambda_q=\diag(\lambda_{q,0},\lambda_{q,1},$
$\cdots,\lambda_{q,N-1})$. Hence,
$$
\det(P_I\bb C_q+\sigma_W^2\bb
I)=\prod_{i=0}^{N-1}(P_I\lambda_{q,i}+\sigma_W^2) \quad\textrm{and}
\quad(P_I\bb C_q+\sigma_W^2\bb I)^{-1}=\bb V_q (P_I
\bb\Lambda_q+\sigma_W^2\bb I)^{-1}\bb V_q^\dag
$$
The ML estimate of $P_I$ is obtained by substituting the above two
relations into $p(\bb Y_q;P_I,{\cal H}_1)$ and finding it maximum.
Moreover, we should also note that $P_I$ is non-negative. Thus, we
have
\begin{equation}\label{eq:snake0908}
\hat{P}_I=\max\left(0,\arg\min_{P}\sum_{i=0}^{N-1}
\left(\ln(P\lambda_{q,i}+\sigma_W^2)+\frac{|\bb v_{q,i}^\dag \bb
Y_q|^2}{P\lambda_{q,i}+\sigma_W^2}\right)\right).
\end{equation}
The general solution of the optimization problem in
(\ref{eq:snake0908}) is unknown, and numerical methods are normally
required. Even if we can solve (\ref{eq:snake0908}), the statistic
distribution of the detector in (\ref{eq:horse0908_1}) is
intractable, which yields threshold determination of the detector
very difficult. On the other hand, under ${\cal H}_0$, the random
variable governing the statistics of $\hat{P}_I$ is zero half of the
time and Gaussian for the other half\footnote{It is known that, if
the probability density function $p(\bb x;\theta)$ of the
observation $\bb x$ satisfies some "regularity" conditions, then the
ML estimate of an unknown parameter $\theta$ is unbiased and
asymptotically Gaussian (see e.g. [\citenum{ref14}, Theorem 7.1]).
Thus, when $N$ is large, the ML estimate in the second argument of
$\max(\cdot,\cdot)$ in (\ref{eq:snake0908}), denoted by
$\tilde{P}_I$, is Gaussian. Since $P_I=0$ under ${\cal H}_0$,
$\tilde{P}_I$ has zero mean and is larger and smaller than 0 with
equal probabilities. It follows that, when $N$ is large, $\hat{P}_I$
is zero half of the time and Gaussian for the other half.}. This is
in contrast to the usual Gaussian asymptotic statistics of an ML
estimate. Thus, the asymptotic chi-squared distribution of GLRT when
$N\rightarrow\infty$ does not hold for (\ref{eq:horse0908_1})
\cite{chernoff54}.

Due to the difficulties encountered by GLRT stated in the previous
paragraph, we resort to a locally most powerful (LMP) detector
[\citenum{ref13},\citenum{poor94}]. We rewrite (\ref{eq:mouse0611})
as $\bb Y_q\sim {\cal CN}(\bb 0, P_I\bb C_q+\sigma_W^2\bb I)$ with
$$
{\cal H}_0: P_I=0 \quad\textrm{versus}\quad {\cal H}_1: P_I>0.
$$
The LMP detector, given by
\begin{equation}\label{eq:tiger0611}
\left.\dfrac{\partial \ln p(\bb Y_q;P_I)}{\partial
P_I}\right|_{P_I=0}=-\sigma_W^{-2}\tr(\bb C_q)+\sigma_W^{-4}\bb
Y_q^\dag\bb C_q\bb Y_q,
\end{equation}
is optimal when $P_I$ is small. Thus, the detector is
\begin{equation}\label{eq:rabbit0611}
T_A(\bb Y_q)=\bb Y_q^\dag\bb C_q\bb
Y_q\mathop{\mathop{\gtrless}_{{\cal H}_0}^{{\cal H}_1}}\gamma_q
\end{equation}
as the remaining part of (\ref{eq:tiger0611}) can be absorbed into
the threshold, where the subscript of $T_A(\cdot)$ indicates it is
for Case A. It is seen that LMP has an advantage that no estimate
for $P_I$ is needed. Moreover, it is almost optimal in the low
signal-to-noise ratio (SNR) region for which signal detection is
inherently a difficult problem. For large departure of $P_I$ from 0,
there is no guarantee of LMP's optimality, and a GLRT would perform
better. However, due to the large SNR, the LMP detector can
generally satisfy the system requirement with the advantage of lower
complexity. An interesting interpretation of LMP detectors as
covariance sequence correlators can be found in [\citenum{poor94},
pp. 80].

Denote by $T_A(\bb Y_q)|_{{\cal H}_i}$ the shorthand for $T_A(\bb
Y_q)$ under ${\cal H}_i$. Let $\bb
W_q=[W_q(0),W_q(1),\cdots,W_q(N-1)]^T$. Under ${\cal H}_0$, elements
of $\bb Y_q=\bb W_q$ are independent, and
$$
T_A(\bb Y_q)|_{{\cal H}_0}=\bb W_q^\dag\bb C_q\bb
W_q=\sum_{i=0}^{N-1}\lambda_{q,i}|\bb v_{q,i}^\dag\bb W_q|^2
$$
is a weighted sum of independent chi-squared random
variables. No general closed form is known for its distribution
[\citenum{poor94}, pp. 74--75]. Thus, we look for its asymptotic
distribution. We have
\begin{equation}\label{eq:mouse0707}
\left.\dfrac{\partial \ln p(\bb Y_q;P_I)}{\partial
P_I}\right|_{P_I=0}=\sum_{n=0}^{N-1} \left.\dfrac{\partial \ln p(
Y_q(n);P_I)}{\partial P_I}\right|_{P_I=0},\quad\textrm{under }{\cal
H}_0,
\end{equation}
which by central limit theorem becomes Gaussian.
Thus,
\begin{equation}\label{eq:cow0707}
T_A(\bb Y_q)|_{{\cal H}_0}\mathop{\sim}^a {\cal N}\left(\sigma_W^2
N,\sigma_W^4\tr(\bb C_q^2)\right),
\end{equation}
where $\mathop{\sim}^a$ indicates the sense of asymptote, and the
formula of
\begin{equation}\label{eq:mouse0821}
\textrm{E}\{\bb x^\dag\bb A\bb x\bb x^\dag\bb
B\bb x\}=\tr(\bb A\bb C)\tr(\bb B\bb C)+\tr(\bb A\bb C\bb B\bb C)
\end{equation}
for $\bb x\sim{\cal CN}(\bb 0,\bb C)$ and Hermitian matrices $\bb A$
and $\bb B$ \cite{miller74} is employed. Under ${\cal H}_1$, due to
the PU signal, elements of $\bb Y_q$ may not be independent. This
makes the distribution of $T_A(\bb Y_q)|_{{\cal H}_1}$ difficult to
analyze. In Appendix~\ref{appendix:3}, the asymptotic distribution
of $T_A(\bb Y_q)|_{{\cal H}_1}$ is examined using the central limit
theorem of an $m$-dependent sequence.

To determine the threshold $\gamma_q$ in (\ref{eq:rabbit0611}), we
represent the cumulative distribution function (CDF) of $T_A(\bb
Y_q)|_{{\cal H}_0}$ as $\textrm{CDF}(x)$. By (\ref{eq:tiger0707}),
$\gamma_q$ is given by
$$
\gamma_q=\textrm{CDF}^{-1}((1-\alpha)^{1/B_\textrm{PU}}),
$$
with an overall false alarm probability $\alpha$. If $N$ is large
enough such that the asymptotic distribution (\ref{eq:cow0707})
holds, $\gamma_q$ can be further written as
$$
\gamma_q=\sigma_W^2 N+\sigma_W^2\sqrt{\tr(\bb C_q^2)}\cdot
Q^{-1}(1-(1-\alpha)^{1/B_\textrm{PU}}),
$$
where $Q(x)$ is the Gaussian right-tail probability. On the other
hand, if the asymptotic distribution of $T_A(\bb Y_q)|_{{\cal H}_0}$
is not valid, histograms of $T_A(\bb Y_q)|_{{\cal H}_0}$ can be
obtained by simulations to get an estimate of $\textrm{CDF}(x)$. We
can simply produce the histogram with $\sigma_W^2=1$. For any
particular $\sigma_W^2$, the corresponding histogram can be easily
mapped from that of $\sigma_W^2=1$.

\subsection{Case B: Known PU Signal Frequency
Band}\label{subsection:cow0420}

In this case, we employ the fact that the PU signal, if present,
appears simultaneously at the sub-carriers from $q_0$ to $q_1$. The
algorithm developed for this case works for PU signal with the
bandwidth $B_\textrm{PU}=q_1-q_0+1\geq 2$. When $B_\textrm{PU}=1$,
the PU signal can be detected by first estimating its received power
and then employing an energy detector
\cite{urkowitz67,kostylev02,digham03,green05,ghasemi05,cabric06,mishra06}.

\begin{figure}
\begin{center}
\begin{tabular}{c}
\psfig{figure=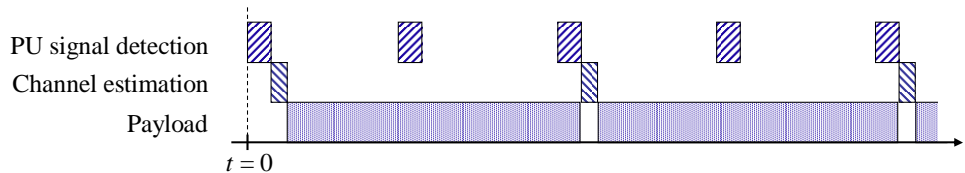,height=2.5cm}
\end{tabular}
\caption{The signalling of the cognitive OFDM system for the
detection algorithm developed in Case B.} \label{fig:fig1}
\end{center}
\end{figure}

Unlike in Case A that the detection is performed when the cognitive
OFDM is not transmitting signals, the detection method presented for
Case B can work when the CU signal is present. The OFDM system
signalling for the detection algorithm in Case B is illustrated in
Fig.~\ref{fig:fig1}. Before initiating ($t=0$), the system performs
a PU signal test at the suspect sub-carriers. If the PU signal is
present, the cognitive system does not send any signal over these
sub-carriers. On the contrary, if the PU signal is absent, channel
estimation of the cognitive system is carried out, and the payloads
are then transmitted over them. If necessary, during payload
transmission, PU signal testing may be executed periodically to
ensure a quick response to the appearance of the primary network. As
shown in Fig.~\ref{fig:fig1}, PU signal detection, channel
estimation and payload transmission are repeated over and over again
(adding PU signal test during payload transmission, if necessary)
until the presence of PU signal is detected. Once PU signal is
detected either at the system initialization or in the middle of a
normal operation, the OFDM system stops transmitting signals over
sub-carriers $q\in[q_0,q_1]$. That is, channel estimation and
payload transmission are suspended, while the PU signal detection is
still performed periodically.

The following situation may arise when PU signal monitoring is done
concurrently with the transmission of the cognitive system. When a
missed detection occurs, the channel estimation will be executed
under the presence of the PU signal, resulting in a poor estimation
accuracy. These inaccurate channel estimates are subsequently
adopted in the next PU signal detection (see
(\ref{eq:rabbit0614})--(\ref{eq:rabbit0621}) below), which is
expected to deteriorate the detection performance. To escape from
the vicious cycle, it is required that PU signal detection is
carried out in the absence of the CU signal. This can be done by
enforcing a "silent period" every a given time interval, during
which the CU should suspend the transmission. 
Although the insertion of silent periods enables the CU to escape
from the vicious cycle, a side effect occurs that the efficiency of
CU is reduced. A discussion of how often and for how long the
cognitive system should remain silent in a given transmission
interval is an important research topic, e.g.
\cite{cordeiro06,ieee802.22,weiss04,kim08}. A trade-off should be
made between two opposing issues of efficiency and integrity of
sensing results to produce a desirable balance defined by a suitable
objective function. For example, in \cite{kim08}, an MAC-layer
sensing period adaptation algorithm is designed to maximize the
discovery of spectrum opportunities.

As described in the previous paragraph, PU signal detection may be
executed when the cognitive OFDM is either on or off. The received
signal at the $q$-th sub-carrier $Y_q(n)$ is given by
(\ref{eq:mouse0530}) with $H_q(n)$ and $S_q(n)$ set to zero when the
OFDM system is not transmitting. For $q_0\leq q\leq q_1$, we build
an observation $\overline{Z}(q)$ from $\{Y_q(n)\}_{n=0}^{N-1}$ such
that the PU signal detection is based on
the observation along the frequency domain, i.e. $\overline{\bb
Z}=[\overline{Z}(q_0),\overline{Z}(q_0+1),\cdots,\overline{Z}(q_1)]^T$.
We choose
\begin{equation}\label{eq:mouse0614}
\overline{Z}(q)=\dfrac{1}{N}\bb Y_q^\dag\bb Y_q,\quad q_0\leq q\leq
q_1,
\end{equation}
because $\overline{Z}(q)$ is the periodogram of the received signal
at the $q$-th sub-carrier averaged over $N$ OFDM symbols. It is well
known that the periodogram is an estimate of the true spectrum of a
signal. Another interpretation of (\ref{eq:mouse0614}) is that the
normalized covariance matrix $\bb C_q$ in the LMP detector of
(\ref{eq:rabbit0611}) is replaced with the identity matrix due to
the unavailability of it. That is, elements of $\bb Y_q$ are
regarded as uncorrelated.

Expanding (\ref{eq:mouse0614}), we obtain
\begin{eqnarray}\label{eq:rabbit0614}
\overline{Z}(q)&=&\frac{1}{N}\sum_{n=0}^{N-1}\left(
|H_q(n)|^2+|I_q(n)|^2+|W_q(n)|^2+\right.\nn\\
&& \left.
2\Re\{H_q(n)S_q(n)I_q^*(n)\}+2\Re\{H_q(n)S_q(n)W_q^*(n)\}+2\Re\{I_q(n)W_q^*(n)\}
\right).
\end{eqnarray}
Depending on whether the OFDM system is transmitting or not,
$H_q(n)$ is either known from channel estimation or equal to zero.
We define
\begin{equation}\label{eq:tiger0621}
m(q):=\frac{1}{N}\sum_{n=0}^{N-1} |\hat{H}_q(n)|^2+\hat{\sigma_W^2},
\end{equation}
where $\hat{H}_q(n)$ and $\hat{\sigma^2_W}$ are the estimates of $H_q(n)$ and $\sigma_W^2$, respectively.
A new observation is built as
\begin{equation}\label{eq:rabbit0621}
Z(q)=\overline{Z}(q)-m(q),\quad q_0\leq q\leq q_1,
\end{equation}
which corresponds to subtracting the first and third terms inside
the brackets of (\ref{eq:rabbit0614}).

Let $\bb Z=[Z(q_0),Z(q_0+1),\cdots,Z(q_1)]^T$ be the observation of
the spectrum sensing problem. It is seen that each component of $\bb
Z$ is involved with a number of contributions from the PU signal, CU
signal, AWGN, estimation errors of $H_q(n)$ and $\sigma_W^2$, and so
on. Thus, it is hard to make a precise statistical description for
$\bb Z$, leading to the difficulty in formulating the corresponding
hypothesis test. To alleviate the problem, here we adopt the model
investigated in \cite{desai03}, where there exists an interfering
signal lying in an arbitrary unknown subspace of the observation
space. Specifically, when the PU signal is present, the observation
vector $\bb Z$ is formulated as
\begin{eqnarray}
\bb Z&=&\bb I+\bb R+\bb N, \label{eq:dragon0409}\\
&=&\bb H\pmb \mu+\bb U\pmb\psi+\sigma^2\overline{\bb
N},\label{eq:dragon0416}
\end{eqnarray}
where
\begin{equation}
\bb I=\frac{1}{N}\left[\sum_{n=0}^{N-1}|I_{q_0}(n)|^2,\sum_{n=0}^{N-1}|I_{q_0+1}(n)|^2,
\cdots,\sum_{n=0}^{N-1}|I_{q_1}(n)|^2\right]^T
\end{equation}
is the contribution of the PU signal, $\bb R$ is the component  due
to an unknown interference, and $\bb N$ is the white noise. The PU
signal $\bb I$ resides in an $(r+1)$-dimensional subspace spanned by
the columns of the known $B_\textrm{PU}\times (r+1)$ matrix $\bb H$,
given as
\begin{equation}\label{eq:mouse0417}
\bb H=[\bb h_0,\bb h_1,\cdots,\bb h_r]=\left(
\begin{array}{cccc}
h_0(0) & h_1(0) & \cdots & h_r(0)\\
h_0(1) & h_1(1) & \cdots & h_r(1)\\
\vdots & \vdots & \ddots & \vdots \\
h_0(q_1-q_0) & h_1(q_1-q_0) & \cdots & h_r(q_1-q_0)
\end{array}
\right),
\end{equation}
and has an unknown gain vector $\pmb\mu$. That is, the powers of the
PU signal across sub-carriers $q\in[q_0,q_1]$ is modeled as a linear
combination of vectors $\{\bb h_i:0\leq i\leq r\}$. The white noise
$\bb N$ is a $B_\textrm{PU}$-dimensional Gaussian random vector
modeled as $\sigma^2\overline{\bb N}$, where the scalar $\sigma^2$
is unknown, and the covariance matrix of $\overline{\bb N}$ is the
identity matrix.
Finally, the vector $\bb R=\bb U\pmb\psi$ accounts for the effects
that are ignored by $\bb I$ and $\bb N$ in (\ref{eq:dragon0409});
both the matrix $\bb U$, whose columns constitute the subspace of
$\bb R$, and the gain vector $\pmb\psi$ are unknown.

It is argued in \cite{desai03} that we require to robustly choose
the unknown matrix $\bb U$ in the formulation of a hypothesis test
such that an adequate level of protection to false alarm as well as
a sufficient detection sensitivity to the signal are both
maintained. Let $\bb G$ be a $B_\textrm{PU}\times
(B_\textrm{PU}-r-1)$ matrix whose columns span the orthogonal
complement of the space generated by $\bb H$, i.e. $\bb G=\bb
H^\perp$. A minimax-based reasoning in [\citenum{desai03},
Appendix~A] leads to explicit and different choices for the unknown
subspace $\bb U$ in the two hypotheses. That is, $\bb U=\bb G$ for
${\cal H}_0$, and $\bb U$ is the zero matrix for ${\cal H}_1$. Thus,
the hypothesis test is
\begin{equation}\label{eq:cow0417}
\begin{array}{ll}
{\cal H}_0: & \bb Z=\bb G\pmb\psi+\sigma_0^2\overline{\bb N},\\
{\cal H}_1: & \bb Z=\bb H\pmb\mu+\sigma_1^2\overline{\bb N},\quad
\bb H\pmb\mu \succcurlyeq\bb 0,
\end{array}
\end{equation}
where $\bb H\pmb\mu\succcurlyeq\bb 0$ means that all elements in
$\bb H\pmb\mu$ are non-negative, and $\pmb\psi$, $\pmb\mu$,
$\sigma_0^2$ and $\sigma_1^2$ are all unknown.
To perform PU signal detection, GLRT of
\begin{equation}\label{eq:dragon0615}
L_G(\bb Z)=\dfrac{p(\bb Z;\hat{\pmb\mu},\hat{\sigma_1^2},{\cal
H}_1)}{p(\bb Z;\hat{\pmb\psi},\hat{\sigma_0^2},{\cal H}_0)}
\end{equation}
is employed, where $\hat{\pmb\mu}$, $\hat{\pmb\psi}$, and
$\hat{\sigma_i^2}$ are ML estimates of $\pmb\mu$, $\pmb\psi$ and
$\sigma^2_i$, respectively. Although, when ${\cal H}_1$ is true,
$\pmb\mu$ and $\sigma_1^2$ are both parameterized by the PU signal;
however, joint estimate of these two unknowns results in a complex
detector structure. Thus, in spite of their dependence, $\pmb\mu$
and $\sigma_1^2$ are estimated separately. The specification of the
PU signal subspace $\bb H$ is only an approximation, and the
performance of the detector depends on whether the linear subspace
spanned by $\{\bb h_i\}$ gives good description of the signal class.
In the case where the correlation matrix of the PU signal is known,
$\bb H$ could be selected as orthogonal eigenvectors of the
correlation matrix, and the subspace dimension $r+1$ can be chosen
based on some information measures \cite{wax85}, e.g. the Akaike
information criterion (AIC) \cite{akaike73} and the minimum
description length (MDL) \cite{rissanen78}. When the correlation
matrix of the PU signal is unknown, the above information measures
and their variations, e.g. \cite{hurvich89}, can assess the
discrepancy between the true and approximating models, which serve
as useful tools in solving the model selection problem. We suppose
that the set of vectors $\{\bb h_i\}$ used to model the PU signal
channel selectivity is suitably chosen. Consequently, $\bb
H\pmb\mu\succcurlyeq\bb 0$ holds most of the time, and the one-sided
test of $\bb H\pmb\mu$ in (\ref{eq:cow0417}) does not bring much
trouble.

It is shown in \cite{desai03} that the likelihood ratio in
(\ref{eq:dragon0615}) for the robust hypothesis test leads to the
matched subspace filter, given by
\begin{equation}\label{eq:mouse0621}
T_B(\bb Z)= \frac{B_\textrm{PU}-r-1}{r+1}\dfrac{\bb Z^T\bb H(\bb
H^T\bb H)^{-1}\bb H^T\bb Z}{\bb Z^T(\bb I-\bb H(\bb H^T\bb
H)^{-1}\bb H^T)\bb Z}.
\end{equation}
It is known that, for the test in (\ref{eq:cow0417}), when $\bb G$
is set as the zero matrix, GLRT yields the matched subspace filter
\cite{scharf94}. Equation (\ref{eq:mouse0621}) demonstrates, even in
the presence of unknown $\bb G\pmb\psi$, the matched subspace
detector is optimal, meaning that it is robust to the interference
whose subspace is unknown.

Under ${\cal H}_0$, the detector is distributed as $T_B(\bb Z)\sim
F_{r+1,B_\textrm{PU}-r-1}$, where $F_{a,b}$ denotes an $F$
distribution with $a$ numerator degrees of freedom and $b$
denominator degrees of freedom. Given threshold $\gamma$, the false
alarm probability is given by
$$
P_\textrm{FA}=Q_{F_{r+1,B_\textrm{PU}-r-1}}(\gamma)
$$
with $Q_{F_{a,b}}(x)$ the right-tail probability of $F_{a,b}$
evaluated at $x$. If $\{\bb h_i\}$ is able to model the PU signal
$\bb I$ well, we have $T_B(\bb Z)\sim
F'_{r+1,B_\textrm{PU}-r-1}(\lambda)$ under ${\cal H}_1$, where
\begin{equation}\label{eq:mouse0420}
\lambda=\dfrac{1}{\sigma_1^2}\sum_{q=q_0}^{q_1}\sum_{n=0}^{N-1}|I_q(n)|^2,
\end{equation}
and $F'_{a,b}(\lambda)$ denotes a noncentral $F$ distribution with
$a$ numerator degrees of freedom, $b$ denominator degrees of freedom
and non-centrality parameter $\lambda$. Thus, the detection
probability with threshold $\gamma$ is
$$
P_\textrm{D}=Q_{F'_{r+1,B_\textrm{PU}-r-1}(\lambda)}(\gamma),
$$
where $Q_{F'_{a,b}(\lambda)}(x)$ is the right-tail probability of
$F'_{a,b}(\lambda)$ evaluated at $x$.

\subsection{Case C: No Prior Knowledge of PU Signal}

In this case, the information of possibly interfered frequency band
is unknown. Consequently, the detection algorithm should be involved
with a search of the interfered band. Intuitively, given the
observation $\bb Z_{0:Q-1}=[Z(0),Z(1),\cdots,Z(Q-1)]^T$, this search
is based on the powers at all sub-carriers, and, if the cognitive
OFDM system is transmitting signals, a sub-carrier with larger
frequency response magnitude $|H_q(n)|$ tends to be judged as the PU
signal is present. To avoid this problem, the search of interfered
band is executed when the cognitive system is not transmitting
signals.

The hypothesis testing associated with Case C is a detection of
abrupt changes \cite{basseville93}, given as
\begin{equation}\label{eq:rabbit0618}
\begin{array}{ll}
{\cal H}_0: & Z(q)\sim U_0(q),\hspace{.8cm}q\in[0,Q-1],  \\
{\cal H}_1: & Z(q)\sim\left\{
\begin{array}{ll}
U_0(q), & q\in[0,q_0-1]\hspace{1mm}\bigcup\hspace{1mm}[q_1+1,Q-1],\\
U_1(q), & q\in[q_0,q_1],
\end{array}
\right.
\end{array}
\end{equation}
where $\{U_0(q)\}_q$ are white Gaussian with variance $\sigma_0^2$,
and $\{U_1(q)\}_{q=q_0}^{q_1}$ are independent Gaussian with the
mean vector modeled by $\{\bb h_i\}$ and variance $\sigma_1^2$. All
of $q_0$, $q_1$, $\sigma_0^2$, $\sigma_1^2$ and the weighting
factors of $\{\bb h_i\}$ are unknown.

We can obtain that the GLRT corresponding to (\ref{eq:rabbit0618})
is
\begin{equation}\label{eq:mouse0619}
\max_{a_0,a_1}\dfrac{(\hat{\sigma_0^2}_{|{\cal H}_0})^{Q/2}}
{(\hat{\sigma_0^2}_{|{\cal
H}_1})^{(Q-a_1+a_0-1)/2}(\hat{\sigma_1^2}_{|{\cal
H}_1})^{(a_1-a_0+1)/2}}
\end{equation}
where
\begin{eqnarray}
\begin{array}{lll}
& & \hat{\sigma_0^2}_{|{\cal H}_0}:=Q^{-1}\bb Z_{0: Q-1}^T\bb Z_{0:
Q-1},\nn\\
& & \hat{\sigma_0^2}_{|{\cal H}_1}:=(Q-a_1+a_0-1)^{-1}(\bb
Z_{0:a_0-1}^T\bb
Z_{0:a_0-1}+\bb Z_{a_1+1:Q-1}^T\bb Z_{a_1+1:Q-1}),\nn\\
\textrm{and} & & \hat{\sigma_1^2}_{|{\cal H}_1}:=(a_1-a_0+1)^{-1}\bb
Z_{a_0:a_1}^T(\bb I-\bb H_{a_0:a_1}(\bb H_{a_0:a_1}^T\bb
H_{a_0:a_1})^{-1}\bb H_{a_0:a_1}^T) \bb Z_{a_0:a_1},\nn
\end{array}
\end{eqnarray}
denote estimate of $\sigma_0^2$ under ${\cal H}_0$, estimate of
$\sigma_0^2$ under ${\cal H}_1$, and estimate of $\sigma_1^2$ under
${\cal H}_1$, respectively, and $\bb H_{a_0:a_1}$ is given in
(\ref{eq:mouse0417}) with $q_0$ and $q_1$ replaced by $a_0$ and
$a_1$, respectively.
Defining $f(a_0,a_1)$ as the target to be maximized in
(\ref{eq:mouse0619}), we consider the false alarm probability for
the detector with threshold $\gamma$, i.e.,
\begin{eqnarray}
P_{\textrm{FA}}&=&\textrm{Prob}\left\{\max_{a_0,a_1}f(a_0,a_1)>\gamma
;{\cal
H}_0 \right\},\nn\\
&=&1-\textrm{Prob}\left\{ f(a_0,a_1)<\gamma,\forall\hspace{1mm}
[a_0,a_1]\subset[0,Q-1];{\cal H}_0 \right\}.\nn
\end{eqnarray}
Since the random variables governing $f(a_0,a_1)$ for different
choices of $a_0$ and $a_1$ are not necessarily independent, the
determination of $P_\textrm{FA}$ and hence the detector threshold
for a given $P_\textrm{FA}=\alpha$ becomes intractable.

To conquer this problem, the PU signal detection is decomposed into
two steps. In the first step, we search for PU signal's frequency
band by $\bb Z_{0:Q-1}$ to get estimates $\hat{q}_0$ and
$\hat{q}_1$. In the second step, we assume $\hat{q}_0$ and
$\hat{q}_1$ obtained in the first step are correct, and we can
consequently perform PU signal detection in the same way as that
proposed for Case B, where PU signal frequency band is known. In
specific, the first step solves the optimization problem of
(\ref{eq:mouse0619}). As the numerator is not a function of $a_0$
and $a_1$, the optimization is equivalent to minimizing the
denominator, i.e.
\begin{equation}\label{eq:cow0416}
(\hat{q}_0,\hat{q_1})=\arg\min_{(a_0,a_1)}(\hat{\sigma_0^2}_{|{\cal
H}_1})^{(Q-a_1+a_0-1)/2}(\hat{\sigma_1^2}_{|{\cal
H}_1})^{(a_1-a_0+1)/2}.
\end{equation}
However, this problem is complex because there are about $Q^2/2$
possible trials for combinations of $a_0$ and $a_1$.

To reduce the computational load, we can simplify the optimization
in (\ref{eq:cow0416}) to one that minimizes the least square (LS)
error between the observations $Z(q)$'s and the estimated PU signal
power, i.e.,
\begin{equation}\label{eq:mouse0620}
(\hat{q}_0,\hat{q}_1)=\arg\min_{(a_0,a_1)}\sum_{q\in[0,Q-1]\setminus[a_0,a_1]}Z(q)^2
+\sum_{q\in[a_0,a_1]}\left(Z(q)-\sum_{i=0}^r \hat{\mu}_{i}
h_i(q-a_0)\right)^2.
\end{equation}
Equation (\ref{eq:mouse0620}) is interpreted as follows. Analogously
to the discussion in Case B, the vector  $\bb
Z_{q_0:q_1}=[Z(q_0),Z(q_0+1),\cdots,Z(q_1)]^T$ contains a signal
lying in the $(r+1)$-dimensional vector space spanned by columns of
$\bb H_{q_0:q_1}$ and having an unknown gain vector $\pmb\mu$.
Consider the second term of the target function at the
right-hand-side of (\ref{eq:mouse0620}). Supposing the PU signal
resides at sub-carriers $q\in[a_0,a_1]$, we find the LS estimate of
the gain vector
$\hat{\pmb\mu}=[\hat{\mu}_0,\hat{\mu}_1,\cdots,\hat{\mu}_r]^T$, i.e.
$\hat{\pmb\mu}=(\bb H_{a_0:a_1}^T\bb H_{a_0:a_1})^{-1}\bb
H_{a_0:a_1}^T\bb Z_{a_0:a_1}$, and compute the LS error between $\bb
Z_{a_0:a_1}$ and $\bb H_{a_0:a_1}\hat{\pmb\mu}$. On the other hand,
in the first term of the target function, only the sum of $Z(q)^2$
is taken into account since sub-carriers
$q\in[0,Q-1]\setminus[a_0,a_1]$ are free from the PU signal.

To compare the ML estimates of $q_0$ and $q_1$ and the suboptimal
ones, the target function of (\ref{eq:mouse0620}) is written as the
vector form
\begin{equation}\label{eq:tiger0416}
\bb Z_{0:a_0-1}^T\bb Z_{0:a_0-1}+\bb Z_{a_1+1:Q-1}^T\bb
Z_{a_1+1:Q-1}+\bb Z_{a_0:a_1}^T(\bb I-\bb H_{a_0:a_1}(\bb
H_{a_0:a_1}^T\bb H_{a_0:a_1})^{-1}\bb H_{a_0:a_1}^T)\bb Z_{a_0:a_1}
\end{equation}
to facilitate examining its relation to the target function of
(\ref{eq:cow0416}), where the third term of (\ref{eq:tiger0416}) is
the LS error yielded by the second term in the target function of
(\ref{eq:mouse0620}).

The solution of (\ref{eq:mouse0620}) can be found by the technique
of dynamic programming (DP) [\citenum{myers81},\citenum{svendsen87}]
as follows. Define
$$
\delta_0(a,b):=\sum_{q\in[a,b]}Z(q)^2\quad\textrm{and}\quad
\delta_1(a,b):=\sum_{q\in[a,b]}\left(Z(q)-\sum_{i=0}^r \hat{\mu}_{i}
h_i(q-a)\right)^2,
$$
where $\{\hat{\mu}_{i}\}_{i=0}^r$ is the LS estimate that minimizes
$\delta_1(a,b)$. Let
\begin{equation}\label{eq:tiger0620}
e(l)=\min_{0\leq a_0\leq l-r} \delta_0(0,a_0-1)+\delta_1(a_0,l),
\quad r\leq l\leq Q-1,
\end{equation}
where the constraint $a_0\leq l-r$ guarantees existence of
$\hat{\mu}_i$'s in $\delta_1(a_0,l)$. The optimization in
(\ref{eq:mouse0620}) is equivalent to
\begin{eqnarray}\label{eq:rabbit0620}
&&\min_{(a_0,a_1)}\delta_0(0,a_0-1)+\delta_1(a_0,a_1)+\delta_0(a_1+1,Q-1)\nn\\
&=&\min_{a_1}\left\{\left(\min_{a_0}\delta_0(0,a_0-1)+
\delta_1(a_0,a_1)\right)+\delta_0(a_1+1,Q-1)\right\}\nn\\
&=&\min_{r\leq a_1\leq Q-1}e(a_1)+\delta_0(a_1+1,Q-1).
\end{eqnarray}
Thus, $\hat{q}_1$ can be found by searching for $a_1$ that minimizes
(\ref{eq:rabbit0620}), and $\hat{q}_0$ is equal to the value of the
argument $a_0$ in (\ref{eq:tiger0620}) that minimizes
$e(\hat{q}_1)$. Note that, the computation of $\delta_1(a_0,l)$'s in
solving (\ref{eq:tiger0620}) can be done recursively by sequential
LS formulas [\citenum{ref14}, pp. 242--251], and the DP can reduce
the complexity of search from the order of $Q^2$ to the order of
$Q$.

Suppose that DP yields correct values of $q_0$ and $q_1$. We then
employ the detector proposed for Case B to decide whether a PU
signal is present in the estimated frequency band.
The performance analysis of the detector is executed as follows. The
false alarm probability $P_\textrm{FA}$ of the detector in Case C is
\begin{equation}\label{eq:rabbit0416}
\sum_{[a_0,a_1]\subset[0,Q-1]}\textrm{Prob}\left\{\hat{q}_0=a_0,\hat{q}_1=a_1;{\cal
H}_0\right\}\cdot \textrm{Prob}\left\{T_B(\bb
Z_{a_0:a_1})>\gamma_{a_0,a_1};{\cal H}_0\right\},
\end{equation}
where the first probability is the one that DP yields the result of
$(\hat{q}_0,\hat{q}_1)=(a_0,a_1)$ under ${\cal H}_0$, and
$T_B(\cdot)$ is given in (\ref{eq:mouse0621}) with $B_\textrm{PU}$
and $\bb H$ there replaced by $a_1-a_0+1$ and $\bb H_{a_0:a_1}$,
respectively. We cannot determine the first probability in
(\ref{eq:rabbit0416}). However, for each DP searching result
$(\hat{q}_0,\hat{q}_1)=(a_0,a_1)$, we can choose a threshold
$\gamma_{a_0,a_1}$ for $T_B(\bb Z_{a_0:a_1})$ such that the second
probability in (\ref{eq:rabbit0416}) is a constant. In this case,
the false alarm probability $P_\textrm{FA}$ becomes
$$
P_\textrm{FA}=\textrm{Prob}\left\{T_B(\bb
Z_{a_0:a_1})>\gamma_{a_0,a_1};{\cal H}_0\right\}.
$$
Given a constraint of $P_\textrm{FA}=\alpha$, we choose
$$
\gamma_{a_0,a_1}=Q^{-1}_{F_{r+1,a_1-a_0-r}}(\alpha),\qquad[a_0,a_1]\subset[0,Q-1].
$$
Detection occurs when, under ${\cal H}_1$, DP returns a correct
result and the decision statistic is greater than the threshold. Let
$p$ denote the probability that the frequency band search returns a
correct result. When the PU signal is well modeled by $\{\bb h_i\}$,
the detection probability $P_\textrm{D}$ is equal to
\begin{eqnarray}
&&p\cdot\textrm{Prob}\left\{T_B(\bb
Z_{q_0:q_1})>Q^{-1}_{F_{r+1,q_1-q_0-r}}(\alpha);{\cal
H}_1\right\}\nn\\
&=&p\cdot
Q_{F'_{r+1,q_1-q_0-r}(\lambda)}\left(Q^{-1}_{F_{r+1,q_1-q_0-r}}(\alpha)\right),\nn
\end{eqnarray}
where $\lambda$ is given in (\ref{eq:mouse0420}).

\section{Simulation Results}

Throughout the simulations, the tonal model presented in
Paragraph~\ref{subsection:2.1} is adopted for the PU signal. The
parameters of the PU and cognitive OFDM systems are $T_s=312.5$~ns,
$T_i=26.6$~$\mu$s, $f_s=3.1$~GHz, $f_i=3.36$~GHz, and $Q=128$. The
number of complex sinusoids $K$ contained in the PU signal is
adjusted according to the PU signal bandwidth $B_\textrm{PU}$.

\begin{figure}
\begin{center}
\begin{tabular}{c}
\psfig{figure=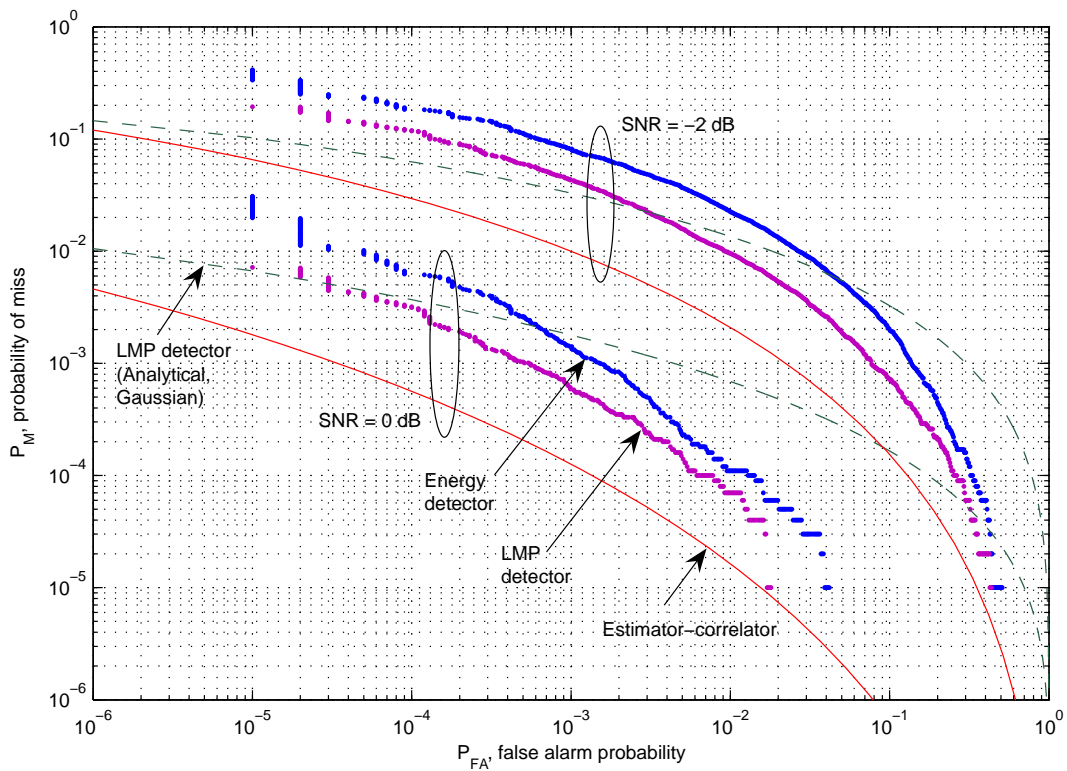,height=11cm}
\end{tabular}
\caption{ROC curves for energy detector, LMP detector and
estimator-correlator performed at a single sub-carrier; $Q=128$,
$N=80$, quiet cognitive system.} \label{fig:CaseA}
\end{center}
\end{figure}

In Fig.~\ref{fig:CaseA}, the receiver operating characteristic (ROC)
of several detectors are shown to illustrate the performance of the
detector proposed in Case A. The probabilities of miss and false
alarm are shown in the vertical and horizontal axes, respectively,
where the detection is performed at a single sub-carrier. The
overall detection performance considering all sub-carriers can be
obtained by (\ref{eq:snake0611}).
The observation length $N$ is set to $80$ OFDM symbols. During the
detection, the cognitive system is not transmitting signals. The
curves in the figure are divided into two groups for the power ratio
of the PU signal and AWGN as $0$ and $-2$ dB. Within each group,
there are four curves. The three solid ones from top to bottom are
simulation result of energy detector (test statistic $\bb
Y_q^\dag\bb Y_q$), simulation result of LMP detector, and analytical
result of estimator-correlator\footnote{The estimator-correlator is
derived from the likelihood ratio test with known received power of
the PU signal. The estimator-correlator and its performance can be
found in [\citenum{ref13}, pp.142].}, respectively; the dashed line
is the analytical result of LMP yielded by Gaussian approximation,
i.e. $T_A(\bb Y_q)|_{{\cal H}_0}$ and $T_A(\bb Y_q)|_{{\cal H}_1}$
are distributed as (\ref{eq:cow0707}) and (\ref{eq:mouse0416}),
respectively. Consistently with our intuition, the
estimator-correlator has the best performance due to its full
knowledge of the observation's PDF, and the energy detector is the
worst since the correlation in the PU signal is not exploited. In
getting the distribution of (\ref{eq:cow0707}), the central limit
theorem for independent and identically distributed random variables
is adopted, while a central limit theorem for $m$-dependent random
variables is used to arrive at the distribution in
(\ref{eq:mouse0416}). The significant discrepancy in the simulated
and analytical results of LMP detector demonstrates the assumptions
in achieving the Gaussian approximations, in particular for
(\ref{eq:mouse0416}), are not valid under the simulation
environments.

\begin{figure}
\begin{center}
\begin{tabular}{c}
\psfig{figure=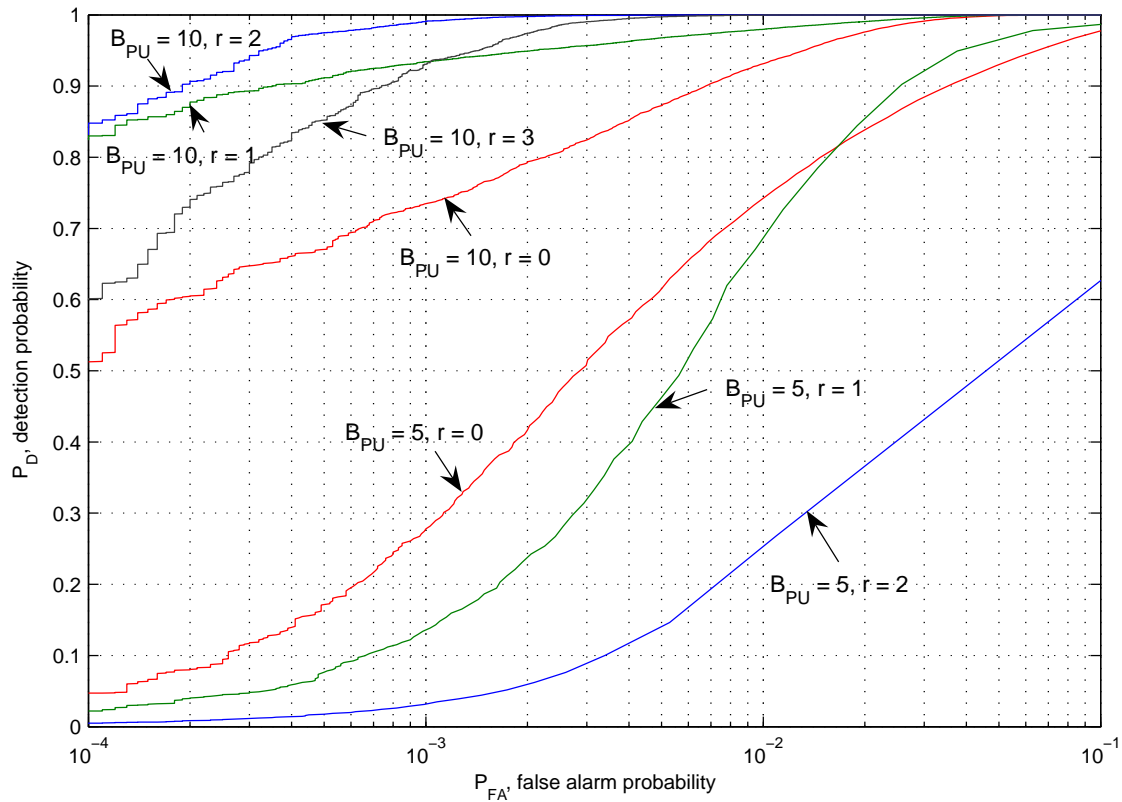,height=11cm}
\end{tabular}
\caption{ROC curves for detector proposed in Case B when the PU
signal experiences a multipath channel with path number equal to
eight and uniform power delay profile; $Q=128$, $N=70$, quiet
cognitive system, average PU signal to AWGN power ratio 0 dB.}
\label{fig:CaseB_1}
\end{center}
\end{figure}

\begin{figure}
\begin{center}
\begin{tabular}{c}
\psfig{figure=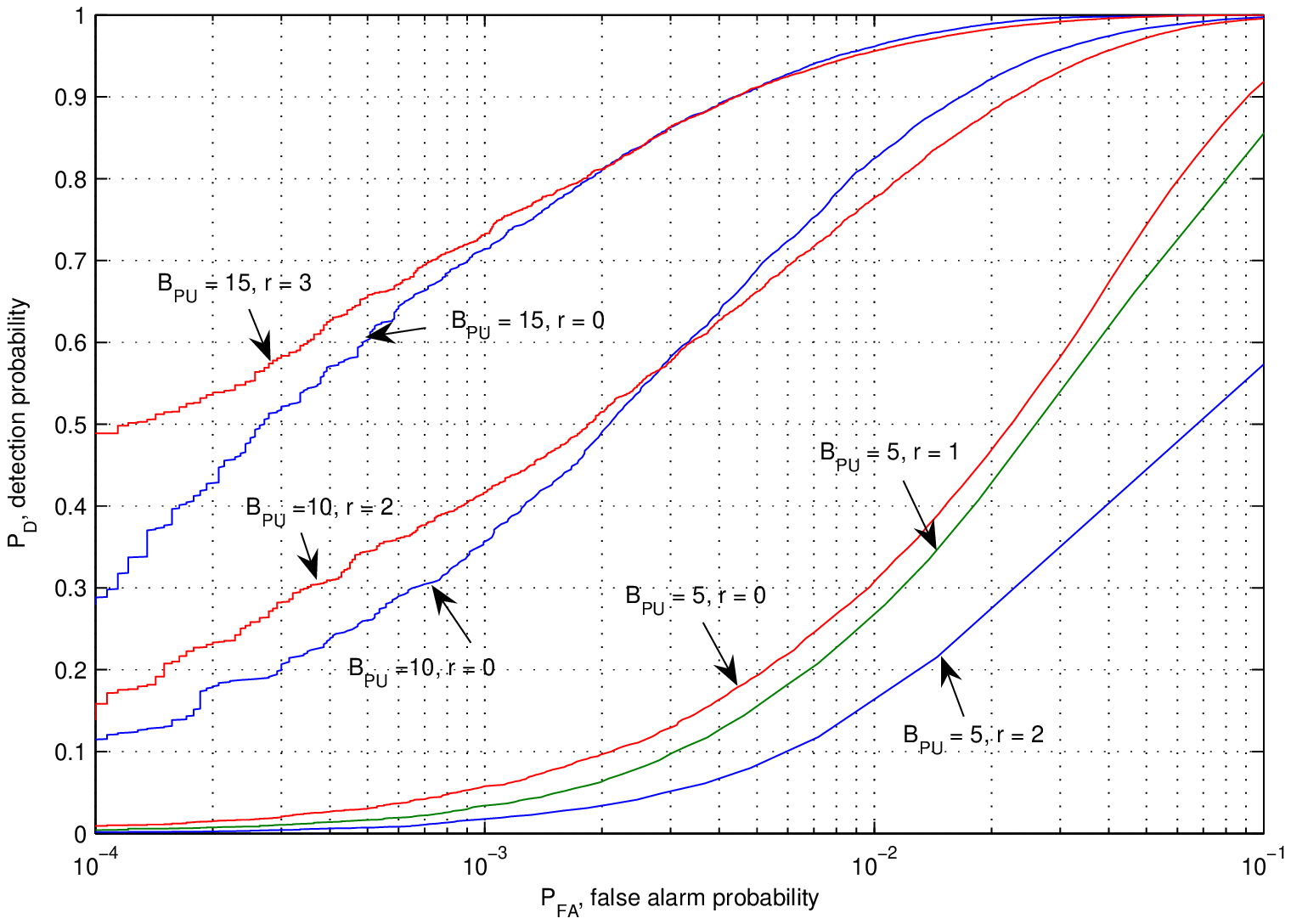,height=11cm}
\end{tabular}
\caption{ROC curves for the detector proposed in Case B when the PU
signal experiences an IEEE 802.15.3a UWB CM3 channel; $Q=128$,
$N=70$, quiet cognitive system, average PU signal to AWGN power
ratio 0 dB.} \label{fig:CaseB_2}
\end{center}
\end{figure}

In Figs.~\ref{fig:CaseB_1} and \ref{fig:CaseB_2}, the simulated ROC
curves of the detector proposed in Case B, i.e.
(\ref{eq:mouse0621}), are plotted for the environments that the PU
signal experiences a channel of eight multipaths with uniform power
delay profile and IEEE 802.15.3a ultrawide band (UWB) CM3 model,
respectively. In either case, a number of channel realizations are
run to obtain an averaged performance. The function $h_i(n)$ in
(\ref{eq:mouse0417}) is set as $n^i$. That is, an $r$-th order
polynomial is used to model PU signal powers across the
sub-carriers. The observation length is $70$ OFDM symbols. During
detection, the cognitive system is quiet.
The average power ratio of the received PU signal and AWGN over the
affected sub-carriers is controlled to be $0$ dB. The bandwidth of
the PU signal $B_\textrm{PU}$ and the order $r$ of the polynomial
used to model the PU signal powers are indicated on each curve. It
is shown that, given the same values of $B_\textrm{PU}$ and $r$, the
detector performance shown in Fig.~\ref{fig:CaseB_1} is better than
that in Fig.~\ref{fig:CaseB_2}, indicating a severe
frequency-selective channel of the PU signal deteriorates the
performance. This is because the polynomial fails to model the PU
signal powers in a hostile channel. Another observation is that,
under the same channel type, detection performance improves when
$B_\textrm{PU}$ increases; whereas increasing the polynomial order
$r$ is not necessarily helpful for detection. This can be explained
as follows.
The numerator and denominator of $T_B(\bb Z)$ can be seen as
estimates of the PU signal power and noise power, respectively
\cite{scharf94}. The dimensions of the signal and noise subspaces
are $r+1$ and $B_\textrm{PU}-r-1$, respectively. If $r$ is too
large, the signal is overestimated, which increases the false alarm
probability; on the other hand, if $r$ is too small, the probability
of miss is increased. As mentioned in
Section~\ref{subsection:cow0420}, information measures can adopted
to choose a suitable value of $r$.

\begin{figure}
\begin{center}
\begin{tabular}{c}
\psfig{figure=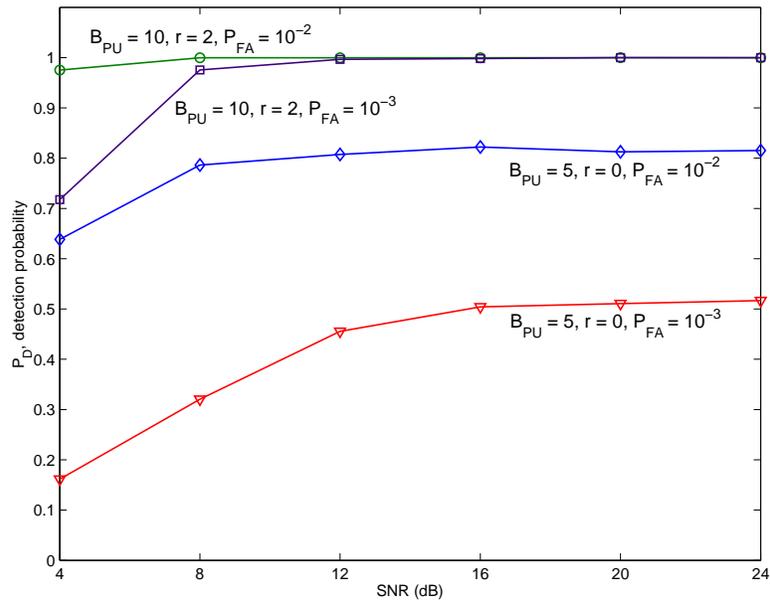,height=8cm}\\
(a) \\
\psfig{figure=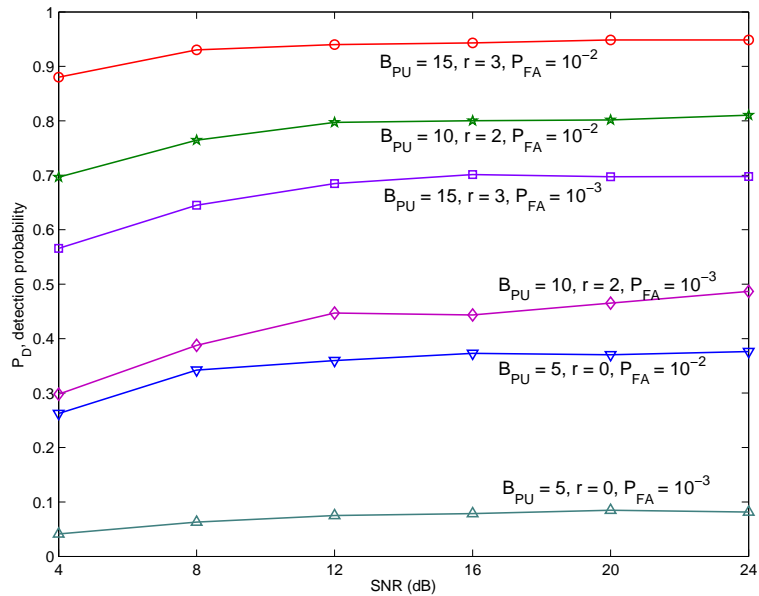,height=8cm}\\
(b)
\end{tabular}
\caption{Detection probability versus the power ratio of the PU
signal to AWGN for the detector proposed in Case B, when the power
ratio of the cognitive signal to AWGN is 8 dB, and the channel
experienced by the PU signal is (a) a multipath channel with path
number equal to eight and uniform power delay profile, (b) an IEEE
802.15.3a UWB CM3 channel; $Q=128$, $N=70$.} \label{fig:CaseB_34}
\end{center}
\end{figure}

In Fig.~\ref{fig:CaseB_34}, the performance of the detector proposed
in Case B is demonstrated when the cognitive system is transmitting
signals. An $r$-th order polynomial is used to model the PU signal
power. The power ratio of the CU signal to AWGN is set to $8$ dB at
every sub-carrier. The observation length is $70$ OFDM symbols. In
Fig.~\ref{fig:CaseB_34}(a), the average PU signal to noise power
ratio versus $P_\textrm{D}$ is plotted when the PU signal
experiences a multipath channel with path number equal to eight and
uniform power delay profile. The bandwidth $B_\textrm{PU}$,
$P_\textrm{FA}$, and polynomial order $r$ are indicated on each
curve. Fig.~\ref{fig:CaseB_34}(b) shows the same information as
Fig.~\ref{fig:CaseB_34}(a) with the channel experienced by the PU
signal being an IEEE 802.15.3a UWB CM3 channel.

\begin{figure}
\begin{center}
\begin{tabular}{c}
\psfig{figure=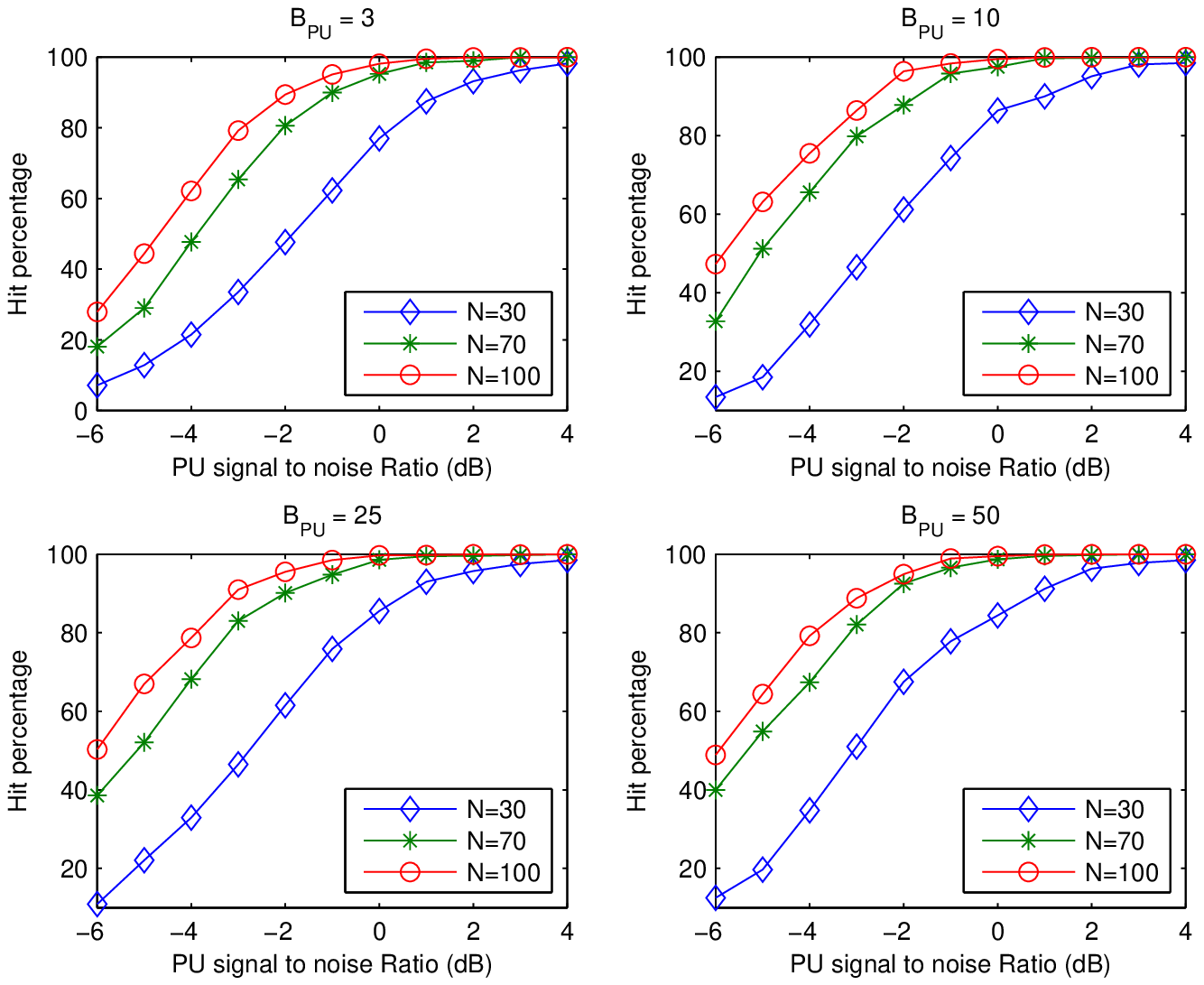,height=11cm}
\end{tabular}
\caption{Hit percentage of the interfered band searching versus the
power ratio of PU signal to AWGN for different values of
$B_\textrm{PU}$; constant PU signal power at all sub-carriers,
$Q=128$, quiet cognitive system.} \label{fig:CaseC}
\end{center}
\end{figure}

In Fig.~\ref{fig:CaseC}, the results of interfered band search based
on (\ref{eq:mouse0620}) of Case C are shown. We try different
$B_\textrm{PU}$ and observation length $N$ with a \textit{constant}
PU signal power over the sub-carriers, i.e. a frequency-flat fading
channel. The dimension of signal subspace is set to $1$, and $\bb
h_0$ is an all-ones vector. A search is regarded to be a hit if
$|q_0-\hat{q}_0|\leq 1$ and $|q_1-\hat{q}_1|\leq 1$. It is shown
that, with an observation length $N=70$, the hit percentage
approaches 100\% at about SNR 1 dB. It is also seen that the hit
percentage is irrelevant to the bandwidth $B_\textrm{PU}$.

\begin{figure}
\begin{center}
\begin{tabular}{c}
\psfig{figure=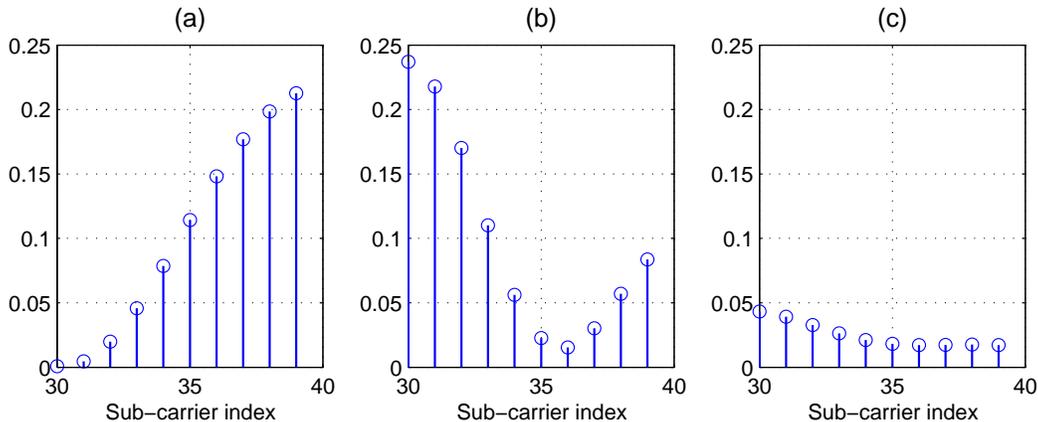,height=6cm,width=14cm}
\end{tabular}
\caption{Examples of frequency-selective fading channels that result
in erroneous estimates of interfered band. The vertical axis is the
squared magnitude of the channel frequency response.}
\label{fig:coop}
\end{center}
\end{figure}

However, when the PU signal experiences severe frequency-selective fading
channel, where there is a notch within the interfered band, our
simulation results indicate that the performance of band search is
not as good as that presented in Fig.~\ref{fig:CaseC}. Examples of
channels that result in erroneous estimates are shown in
Fig.~\ref{fig:coop}, where the responses (sub-carriers $30$ to $39$)
at which the PU signal resides are plotted. For each sub-figure, the
horizontal and vertical axes are sub-carrier index and squared
magnitude of channel frequency response, respectively. In
Figs.~\ref{fig:coop}(a) and \ref{fig:coop}(b), there are spectrum
notches within the band; while, in Fig.~\ref{fig:coop}(c), all the
sub-carriers suffer from deep fades. In the cases of
Figs.~\ref{fig:coop}(a) and \ref{fig:coop}(b), the estimate
$[\hat{q}_0,\hat{q}_1]$ is a subset of the true band $[q_0,q_1]$; in
the case of Fig.~\ref{fig:coop}(c), $[\hat{q}_0,\hat{q}_1]$ is not
even a subset of $[q_0,q_1]$. Such problem arising in a severe
frequency-selective channel can be conquered by cooperation among
CUs, as channels from the PU signal source to CUs of non-proximity
can be regarded as independent.

\section{Conclusion}

In this paper, the problem of PU signal detection in an OFDM based
cognitive system is addressed. We categorize the amount of PU
signal's prior knowledge into three cases. In each case, a
Neyman-Pearson detector that exploits the prior information is
proposed.

In Case A, it is assumed that the PU signal model is known, and the
PDFs of the received signal under both hypotheses are completely
available except for the received power. The frequency band that the
PU signal resides is known as well. Due to the difficulties of
finding the received signal power estimate and the detector
threshold, the use of a GLRT detector is not suggested. In stead, an
LMP detector is employed due to the advantages of being optimal for
weak PU signal and no need to get the unknown power estimate. Since
the covariance matrix and the received signal power are distinct at
every sub-carrier, the detection is performed individually at
sub-carriers, and the final decision is formed by an OR operation of
the results of the individual detections. The simulation result of
an LMP detector is compared with an energy detector (information of
PU signal covariance matrix not exploited) and estimator-correlator
(perfect knowledge of PU signal received power). The performance of
LMP detector is between the other two.

In Case B, we assume the only PU signal prior information is its
residing sub-carriers. The detector proposed for this case exploits
the fact that, once the PU signal appears, it interferers a
consecutive segment of sub-carriers simultaneously. Measurements
obtained at all these sub-carriers are taken as the observation. The
hypothesis test is an unknown subspace signal detection in an
unknown interference and a Gaussian noise with unknown variance. The
result of \cite{desai03} is employed to robustly choose the subspace
where the unknown interference locates, and it is shown a matched
subspace detector is the GLRT of the hypothesis testing.

In Case C, no prior knowledge about the PU signal is available. The
detection is involved with a search of interfered sub-carriers. It
is shown that the GLRT has high complexity and is difficult to find
the output distribution, leading to an undetermined threshold. Thus,
the detection is divided into two steps. The first step searches for
the interfered band using an ML criterion, and the second step
decides whether a PU signal is present in the estimated interfered
band. It is seen that the second step corresponds to the problem
considered in Case B. The first step is further simplified by
employing an LS criterion instead of ML, which enables the use of
the DP technique to solve the optimization problem. Simulation
results show the search of interfered band has a very high accuracy
if the channel experienced by the PU signal is frequency-flat faded.
However, if the channel is severely frequency-selective, the
estimation accuracy degrades. It is believed that cooperation among
CUs is helpful in conquering this problem.

\appendices

\section{Normalized Covariance Matrices of PU Signal Models}

\subsection{Tonal PU Signal}\label{appendix:1}

Let the cognitive OFDM system have symbol duration $T_s$, carrier
frequency $f_s$, and its zero-th symbol start at $t=t_0$. We suppose
that, at the time the $n$-th OFDM symbol is received, i.e.
$nT_s+t_0\leq t< (n+1)T_s+t_0$, it is within the span of PU signal's
$l$-th symbol.\footnote{Although it is possible that an OFDM symbol
crosses the boundary of two PU signal symbols, as $T_i$ is in
general much larger than $T_s$, the contribution resulting from this
situation is negligible. For instance, if Mobile WiMAX and mulriband
(MB)-OFDM-UWB are the sources of the PU signal and the cognitive
OFDM system, respectively, the value of $T_i/T_s$ is as large as
$329$.} When detecting the $n$-th OFDM symbol, at the down-converter
output of the receiver, the contribution resulting from the PU
signal is given by
$$
\sum_{k=0}^{K-1} \zeta_{l,k}X_{l,k}e^{j 2\pi k(t-l
T_i)/T_i}e^{j\{2\pi[f_{i}(t-l T_i)-f_{s}(t-n T_s-t_0)]+\phi\}},\quad
nT_s+t_0\leq t<(n+1)T_s+t_0.
$$
It is then sampled every $T_d:=T_s/Q$ seconds, resulting in $Q$
samples of
\begin{equation}\label{eq:mouse0604}
i_p=
\sum_{k=0}^{K-1} \zeta_{l,k}X_{l,k}e^{j2\pi (\Delta f+k/T_i)p
T_d}e^{j\theta(n,l,k)},\quad 0\leq p\leq Q-1,
\end{equation}
where $\Delta f=f_i-f_s$ and $\theta(n,l,k)=2\pi [(n
T_s+t_0)(f_i+k/T_i)-l f_i T_i]+\phi$. The discrete-time signal
$\{i_p\}$ is passed through a $Q$-point DFT, which gives
\begin{eqnarray}\label{eq:mouse0908}
I_q(n)&=&\sum_{p=0}^{Q-1}i_p e^{-j 2\pi pq/Q}\nn\\
&=&\sum_{k=0}^{K-1} \zeta_{l,k}X_{l,k} e^{j\theta(n,l,k)}
e^{j\pi\beta_{k,q}(Q-1)}\dfrac{\sin(\pi\beta_{k,q}
Q)}{\sin(\pi\beta_{k,q})},\quad 0\leq q\leq Q-1,
\end{eqnarray}
where $\beta_{k,q}=(\Delta f+k/T_i)T_d-q/Q$. It is seen from
(\ref{eq:mouse0908}) that, due to central limit theorem,
$\{I_q(n)\}_{n=0}^{N-1}$ can be approximated by a Gaussian random
sequence when the number of complex sinusoids $K$ is large enough.

Let us denote by event $\cal A$ that the $n$- and $m$-th OFDM
symbols both fall within the span of the $l$-th PU signal symbol.
Clearly, $I_q(n)$ and $I_q(m)$ are zero-mean random variables.
Conditioned on event $\cal A$, the correlation of $I_q(n)$ and
$I_q(m)$ is
$$
\textrm{E}\{I_q(n) I_q^*(m)|{\cal A}\}=
\textrm{E}\{|\zeta_{l,k}|^2\} e^{j 2\pi(n-m)T_s f_i}
\sum_{k=0}^{K-1} e^{j 2\pi(n-m)k T_s
/T_i}\dfrac{\sin^2(\pi\beta_{k,q} Q)}{\sin^2(\pi\beta_{k,q})},
$$
where the expectation at the left-hand-side is with respect to
symbols $X_{l,k}$ of the PU signal and the fading coefficients
$\zeta_{l,k}$. On the contrary, if OFDM symbols $n$ and $m$ fall
within the spans of distinct symbols of the PU signal, we have
$$
\textrm{E}\{I_q(n) I_q^*(m)|\overline{{\cal A}}\}=0,
$$
where $\overline{{\cal A}}$ is the complement of $\cal A$.

Let $\eta:= \lfloor T_i/T_s\rfloor$ with $\lfloor x\rfloor$ the
largest integer no greater than $x$. The probability
$\textrm{Pr}\{{\cal A}\}$ of event $\cal A$ is roughly equal to
$$
\textrm{Pr}\{{\cal A}\}=\left\{
\begin{array}{ll}
1-\dfrac{|n-m|}{\eta}, & |n-m|\leq \eta-1,\\
0, & \textrm{otherwise},
\end{array}
\right.
$$
where we omit the case the OFDM symbol(s) fall at the border of two
PU signal symbols. Thus, the $(n,m)$-th element of the covariance
matrix of $\{I_q(n)\}_{n=0}^{N-1}$ is given by
$$
\textrm{Pr}\{{\cal A}\}\cdot\textrm{E}\{I_q(n)I^*_q(m)|{\cal A}\},
$$
which leads to the normalized covariance matrix $\bb C_q$ in
(\ref{eq:mouse0605}) assuming, for each particular $l$,
$\zeta_{l,k}$'s are identically distributed.

\subsection{AR PU Signal}\label{appendix:2}

The method to obtain the normalized covariance matrix for AR PU
signal basically follows the line presented in [\citenum{wang2004},
pp. 414]. Since $\bb I_q$ has zero mean, the covariance matrix of
$\bb I_q$ is given by
\begin{equation}\label{eq:rabbit0706}
\textrm{E}\{\bb I_q \bb I_q^\dag\}=\bb F_q \bb R_{\bb i} \bb
F_q^\dag,
\end{equation}
where $^\dag$ denotes Hermitian transpose, and $\bb R_{\bb i}$ is
the correlation matrix of the time-domain PU signal signal $\bb i$.
From (\ref{eq:dragon0908}), it is readily seen that
\begin{equation}\label{eq:cow0605}
\left(
\begin{array}{ccccccc}
1 & & & & & & \\
 & \ddots & & & & & \\
 & & 1 & & & & \\
\phi_r & \cdots & \phi_{1} & 1 & & &  \\
 & \phi_r & \cdots & \phi_{1} & 1 & & \\
 & & \ddots & & & \ddots & \\
& & & \phi_r & \cdots & \phi_{1} & 1
\end{array}
\right) \left(
\begin{array}{c}
i_0\\
\vdots\\
i_{r-1}\\
i_{r}\\
i_{r+1}\\
\vdots\\
i_{QN-1}
\end{array}
\right) = \left(
\begin{array}{c}
i_0\\
\vdots\\
i_{r-1}\\
e_{r}\\
e_{r+1}\\
\vdots\\
e_{QN-1}
\end{array}
\right),
\end{equation}
which has a notational form of
\begin{equation}\label{eq:rabbit0908}
\bb A \left(
\begin{array}{c}
\bb i_{0:r-1}\\
\bb i_{r:QN-1}
\end{array}
\right) = \left(
\begin{array}{c}
\bb i_{0:r-1}\\
\bb e_{r:QN-1}
\end{array}
\right),
\end{equation}
where notation definitions can be easily mapped between
(\ref{eq:cow0605}) and (\ref{eq:rabbit0908}). Right-multiplying both
sides of (\ref{eq:rabbit0908}) by their Hermitian transposes and
then taking expectations, we obtain
\begin{equation}\label{eq:cow0606}
\bb A\bb R_{\bb i} \bb A^\dag= \left(
\begin{array}{cc}
\bb R_{\bb i_{0:r-1}} & \bb 0\\
\bb 0 & \nu^2 \bb I
\end{array}
\right),
\end{equation}
where $\bb R_{\bb i_{0:r-1}}=\textrm{E}\{\bb i_{0:r-1}\bb
i_{0:r-1}^\dag\}$ is unknown. Since $\bb A$ is non-singular, we have
\begin{equation}\label{eq:mouse0706}
{\bb R}_{\bb i}^{-1}=\bb A^\dag \left(
\begin{array}{cc}
\bb R_{\bb i_{0:r-1}}^{-1} & \bb 0\\
\bb 0 & \nu^{-2} \bb I
\end{array}
\right) \bb A.
\end{equation}
Partition $\bb A$ into the following four blocks:
\begin{equation}\label{eq:cow0706}
\bb A=\left(
\begin{array}{cc}
\bb I & \bb 0 \\
\bb A_{21} & \bb A_{22}
\end{array}
\right),
\end{equation}
where the dimensions of $\bb A_{21}$ and $\bb A_{22}$ are
$(QN-r)\times r$ and $(QN-r)\times(QN-r)$, respectively. Plugging
(\ref{eq:cow0706}) into (\ref{eq:mouse0706}), we can write
\begin{equation}\label{eq:tiger0706}
{\bb R}_{\bb i}^{-1}=\left(
\begin{array}{ll}
\bb R_{\bb i_{0:r-1}}^{-1}+\nu^{-2}\bb A_{21}^\dag\bb A_{21} & \nu^{-2}\bb A_{21}^\dag\bb A_{22} \\
\nu^{-2}\bb A_{22}^\dag\bb A_{21} & \nu^{-2}\bb A_{22}^\dag\bb
A_{22}
\end{array}
\right).
\end{equation}
It is known that the inverse of a nonsingular Toeplitz matrix is
persymmetric (i.e., symmetric about the northeast-southwest
diagonal) \cite{golub89}. Since ${\bb R}_{\bb i}$ is Toeplitz, $\bb
R_{\bb i_{0:r-1}}$ can be determined by comparing $\nu^{-2}\bb
A_{21}^\dag\bb A_{21}$ and $\nu^{-2}\bb A_{22}^\dag\bb A_{22}$ in
the northwest and southeast blocks of the block matrix in
(\ref{eq:tiger0706}). Then, the right-hand-side of
(\ref{eq:rabbit0706}) can be written as
\begin{equation}\label{eq:dragon0706}
{\bb F}_q {\bb A}^{-1} \left(
\begin{array}{cc}
\bb R_{\bb i_{0:r-1}} & \bb 0\\
\bb 0 & \nu^2\bb I
\end{array}
\right) ({\bb A}^\dag)^{-1} {\bb F}_q^\dag.
\end{equation}
Note that since the parameters $\{\phi_j\}_{j=1}^r$ are obtained
when the PU signal $\{i_p\}$ has unit power, the matrix given in
(\ref{eq:dragon0706}) is the normalized covariance matrix of $\bb
I_q$.

\section{Asymptotic Distribution of LMP Detector Output under ${\cal H}_1$}\label{appendix:3}

To consider the asymptotic distribution of the LMP detector output
under ${\cal H}_1$, we employ a central limit theorem for an
$m$-dependent sequence. The following definition and theorem are
helpful.

\begin{definition}\label{def:1}
[\citenum{ferguson96}, pp. 69] A sequence of random variables,
$X_1,X_2,\cdots,$ is said to be $m$-dependent if for every integer,
$s\geq 1$, the sets of random variables $\{X_1,\cdots,X_s\}$ and
$\{X_{m+s+1},X_{m+s+2},\cdots\}$ are independent. \hfill{\small
$\blacksquare$}
\end{definition}

\begin{theorem}\label{theorem:1}
[\citenum{ferguson96}, pp. 70] Let $X_1,X_2,\cdots,$ be a stationary
$m$-dependent sequence with finite variance and let
$S_n=\sum_{i=1}^n X_i$. Then
$[S_n-\textrm{E}(S_n)]/\sqrt{\textrm{var}(S_n)}$ converges in
distribution to ${\cal N}(0,1)$.\hfill{\small $\blacksquare$}
\end{theorem}

In the following, we will show that the LMP detector output $\bb
Y_q^\dag\bb C_q\bb Y_q$ under ${\cal H}_1$ is a sum of random
variables in an $m$-dependent sequence. We have
$$
\bb Y_q^\dag \bb C_q\bb Y_q=\sum_{i=0}^{N-1}\sum_{j=0}^{N-1}
Y_q(i)^* C_q(i,j)Y_q(j)= \sum_{i=0}^{N-1} Y_q(i)^*\sum_{j=0}^{N-1}
C_q(i,j)Y_q(j),
$$
where $C_q(i,j)$ is the $(i,j)$-th entry of $\bb C_q$. Define a
random variable $X_i$ as
$$
X_i=Y_q(i)^*\sum_{j=0}^{N-1}
C_q(i,j)Y_q(j)=Y_q(i)^*\sum_{j=\max(0,i-\eta+1)}^{\min(i+\eta-1,N-1)}
C_q(i,j)Y_q(j),
$$
where $\eta$ satisfies that $C_q(i,j)\approx 0$ whenever $|i-j|\geq
\eta$. In the case of a tonal PU signal model, we have $\eta=\lfloor
T_i/T_s\rfloor$. It is clear to see $\{X_i\}$ is an $m$-dependent
sequence, and $\bb Y_q^\dag\bb C_q\bb Y_q$ is a sum of $m$-dependent
random variables. By Theorem~\ref{theorem:1}, if $\{X_i\}$ is
stationary and each $X_i$ has finite variance, then the LMP detector
output under ${\cal H}_1$, i.e. $T_A(\bb Y_q)|_{{\cal H}_1}$, is
asymptotically Gaussian. At this moment, we have
\begin{equation}\label{eq:mouse0416}
T_A(\bb Y_q)|_{{\cal H}_1}\mathop{\sim}^a{\cal N}(\textrm{tr}\{\bb
C_q(P_q\bb C_q+\sigma_W^2\bb I)\},\textrm{tr}\{(\bb C_q(P_q\bb
C_q+\sigma_W^2\bb I))^2\}),
\end{equation}
where
the statistical property of Gaussian in (\ref{eq:mouse0821}) is
used.

\bibliography{Hwang}

\begin{thebibliography}{10}

\bibitem{FCC1}
FCC,
\newblock {\em Spectrum Policy Task Force Report},
\newblock ET Docket No. 02-155, Nov. 02, 2002.

\bibitem{mitola99}
J.~Mitola,
\newblock ``{Cognitive radio: making software radio more personal},''
\newblock {\em IEEE Pers. Commun.}, vol. 6, no. 4, pp. 48--52, Aug. 1999.

\bibitem{mitola00}
J.~Mitola,
\newblock {\em Cognitive Radio: An Integrated Agent Architecture for Software
  Defined Radio},
\newblock Ph.D. thesis, Royal Institute of Technology (KTH), Sweden, May 2000.

\bibitem{urkowitz67}
H.~Urkowitz,
\newblock ``{Energy detection of unknown deterministic signals},''
\newblock {\em Proceedings of IEEE}, vol. 55, no. 4, pp. 523--531, Apr. 1967.

\bibitem{kostylev02}
V.~I. Kostylev,
\newblock ``{Energy detection of a signal with random amplitude},''
\newblock in {\em Proc. IEEE ICC}, Apr. 2002, pp. 1606--1610.

\bibitem{digham03}
F.~F. Digham, M.-S. Alouini, and M.~K. Simon,
\newblock ``{On the energy detection of unknown signals over fading
  channels},''
\newblock in {\em Proc. IEEE ICC}, May 2003, pp. 3575--3579.

\bibitem{green05}
Marilynn~P. Wylie-Green,
\newblock ``{Dynamic spectrum sensing by multiband OFDM radio for interference
  mitigation},''
\newblock in {\em Proc. IEEE Int. Symp. on New Frontiers in Dynamic Spectrum
  Access Networks}, Nov. 2005, pp. 619--625.

\bibitem{ghasemi05}
A.~Ghasemi and E.~S. Sousa,
\newblock ``{Collaborative spectrum sensing for opportunistic access in fading
  environments},''
\newblock in {\em Proc. IEEE Int. Symp. on New Frontiers in Dynamic Spectrum
  Access Networks}, Nov. 2005, pp. 131--136.

\bibitem{cabric06}
D.~Cabric, A.~Tkachenko, and R.~W. Brodersen,
\newblock ``{Experimental study of spectrum sensing based on energy detection
  and network cooperation},''
\newblock in {\em Proc. ACM 1st Int. Workshop on Technology and Policy for
  Accessing Spectrum}, Aug. 2006.

\bibitem{mishra06}
S.~M. Mishra, A.~Sahai, and R.~W. Brodersen,
\newblock ``{Cooperative sensing among cognitive radios},''
\newblock in {\em Proc. IEEE Int. Conf. Commun. (ICC) 2006}, June 2006, pp.
  1658--1663.

\bibitem{oner07}
M.~{\"{O}}ner and F.~Jondral,
\newblock ``{On the extraction of the channel allocation information in
  spectrum pooling systems},''
\newblock {\em IEEE J. Select. Areas Commun.}, vol. 25, no. 3, pp. 558--565,
  Apr. 2007.

\bibitem{ghozzi06}
M.~Ghozzi, M.~Dohler, F.~Marx, and J.~Palicot,
\newblock ``{Cognitive radio: methods for the detection of free bands},''
\newblock {\em Comptes Rendus Physique, Elsevier}, vol. 7, pp. 794--804, 2006.

\bibitem{han06}
N.~Han, S.~Shon, J.~H. Chung, and J.~M. Kim,
\newblock ``{Spectral correlation based signal detection method for spectrum
  sensing in IEEE 802.22 WRAN systems},''
\newblock in {\em Proc. Int. Conf. Adv. Commun. Technol.}, Feb. 2006, vol.~3,
  pp. 1765--1770.

\bibitem{lunden07}
J.~Lunden, V.~Koivunen, A.~Huttunen, and H.~V. Poor,
\newblock ``{Spectrum sensing in cognitive radios based on multiple cyclic
  frequencies},''
\newblock in {\em Proc. 2nd Int. Conf. Cognitive Radio Oriented Wireless
  Networks and Communications}, Jul. 31-Aug. 3, 2007.

\bibitem{visotsky05}
E.~Visotsky, S.~Kuffner, and R.~Peterson,
\newblock ``{On collaborative detection of TV transmissions in support of
  dynamic spectrum sensing},''
\newblock in {\em Proc. IEEE Int. Symp. on New Frontiers in Dynamic Spectrum
  Access Networks}, Nov. 2005, pp. 338--345.

\bibitem{weiss03}
T.~Weiss, J.~Hillenbrand, and F.~Jondral,
\newblock ``{A diversity approach for the detection of idle spectral resources
  in spectrum pooling systems},''
\newblock in {\em Proc. of the 48th Int. Sci. Colloquium}, Sept. 2003.

\bibitem{ganesan05}
G.~Ganesan and Y.~G. Li,
\newblock ``{Agility improvement through cooperation diversity in cognitive
  radio},''
\newblock in {\em Proc. IEEE Globecom}, Nov. 2005, vol.~5, pp. 2505--2509.

\bibitem{ganesan07}
G.~Ganesan and Y.~G. Li,
\newblock ``{Cooperative spectrum sensing in cognitive radio: Part I: two user
  networks},''
\newblock {\em IEEE Trans. Wireless Commun.}, vol. 6, no. 6, pp. 2204--2213,
  June 2007.

\bibitem{ganesan07_1}
G.~Ganesan and Y.~G. Li,
\newblock ``{Cooperative spectrum sensing in cognitive radio: Part II:
  multiuser networks},''
\newblock {\em IEEE Trans. Wireless Commun.}, vol. 6, no. 6, pp. 2214--2222,
  June 2007.

\bibitem{gandetto07}
M.~Gandetto and C.~Regazzoni,
\newblock ``{Spectrum sensing: a distributed approach for cognitive
  terminals},''
\newblock {\em IEEE J. Select. Areas Commun.}, vol. 25, no. 3, pp. 546--557,
  Apr. 2007.

\bibitem{cabric04}
D.~Cabric, S.~M. Mishra, and R.~W. Brodersen,
\newblock ``{Implementation issues in spectrum sensing for cognitive radios},''
\newblock in {\em Proc. The Thirty-Eighth Asilomar Conference on Signals,
  Systems and Computers}, Nov. 2004, vol.~1, pp. 772--776.

\bibitem{mishra2007}
S.~M. Mishra, S.~ten Brink, R.~Mahadevappa, and R.~W. Brodersen,
\newblock ``{Detect and avoid: an ultra-wideband/WiMax coexistence
  mechanism},''
\newblock {\em IEEE Communications Magazine}, vol. 45, no. 6, pp. 68--75, June
  2007.

\bibitem{akyildiz06}
I.~F. Akyildiz, W.-Y. Lee, M.~C. Vuran, and S.~Mohanty,
\newblock ``{NeXt generation/dynamic spectrum access/cognitive radio wireless
  networks: A survey},''
\newblock {\em Computer Networks: The Int. Journal of Computer and
  Telecommunications Networking}, vol. 50, no. 13, pp. 2127--2159, Sept. 2006.

\bibitem{tang05}
H.~Tang,
\newblock ``{Some physical layer issues of wide-band cognitive radio
  systems},''
\newblock in {\em Proc. IEEE Int. Symp. on New Frontiers in Dynamic Spectrum
  Access Networks}, Nov. 2005, pp. 151--159.

\bibitem{trees01}
H.~L.~Van Trees,
\newblock {\em Detection, Estimation, and Modulation Theory: Part I. Detection,
  Estimation, and Linear Modulation Theory},
\newblock John Wiley \& Sons, Inc., 2001.

\bibitem{srikanteswara07}
S.~Srikanteswara, Guoqing Li, and C.~Maciocco,
\newblock ``{Cross layer interference mitigation using spectrum sensing},''
\newblock in {\em Proc. 2007 IEEE Global Telecommunications Conference
  (Globecom)}, Nov. 2007, pp. 3553--3557.

\bibitem{ref14}
S.~M. Kay,
\newblock {\em Fundamentals of Statistical Signal Procssing: Estimation
  Theory}, vol.~1,
\newblock Prentice Hall PTR, 1993.

\bibitem{chernoff54}
H.~Chernoff,
\newblock ``{On the distribution of the likelihood ratio},''
\newblock {\em Ann. Math. Statist.}, vol. 25, no. 3, pp. 573--578, Sept. 1954.

\bibitem{ref13}
S.~M. Kay,
\newblock {\em Fundamentals of Statistical Signal Procssing: Detection Theory},
  vol.~2,
\newblock Prentice Hall PTR, 1998.

\bibitem{poor94}
H.~V. Poor,
\newblock {\em An Introduction to Signal Detection and Estimation},
\newblock Springer-Verlag, New York, 2nd edition, 1994.

\bibitem{miller74}
K.~S. Miller,
\newblock {\em Complex Stochastic Processes},
\newblock Addison-Wesley, Reading, Massm, 1974.

\bibitem{cordeiro06}
C.~Cordeiro, K.~Challapali, and M.~Ghosh,
\newblock ``{Cognitive PHY and MAC layers for dynamic spectrum access and
  sharing of TV bands},''
\newblock in {\em Proc. 2006 First Int'l Workshop Technology and Policy for
  Accessing Spectrum (TAPAS '06)}, Aug. 2006.

\bibitem{ieee802.22}
{\em IEEE 802.22 Working Group on Wireless Regional Area Networks},
\newblock http://www.ieee802.org/22/, 2008.

\bibitem{weiss04}
T.~A. Weiss and F.~K. Jondral,
\newblock ``{Spectrum pooling: an innovative strategy for the enhancement of
  spectrum efficiency},''
\newblock {\em IEEE Communications Magazine}, vol. 42, no. 3, pp. 8--14, Mar.
  2004.

\bibitem{kim08}
H.~Kim and K.~G. Shih,
\newblock ``{Efficient discovery of spectrum opportunities with MAC-layer
  sensing in cognitive radio networks},''
\newblock {\em IEEE Trans. on Mobile Computing}, vol. 7, no. 5, pp. 533--545,
  May 2008.

\bibitem{desai03}
M.~N. Desai and R.~S. Mangoubi,
\newblock ``{Robust Gaussian and non-Gaussian matched subspace detection},''
\newblock {\em IEEE Trans. on Signal Processing}, vol. 51, no. 12, pp.
  3115--3127, Dec. 2003.

\bibitem{wax85}
M.~Wax and T.~Kailath,
\newblock ``{Detection of signals by information theoretic criteria},''
\newblock {\em IEEE Trans. Acoust. Speech Signal Process.}, vol. 33, no. 2, pp.
  387--392, Apr. 1985.

\bibitem{akaike73}
H.~Akaike,
\newblock ``{Information theory as an extension of the maximum likelihood
  principle},''
\newblock in {\em B. N. Petrov and F. Csaki (Eds.), Second International
  Symposium on Information Theory, Budapest, Akademiai Kiado}, 1973, pp.
  267--281.

\bibitem{rissanen78}
J.~Rissanen,
\newblock ``{Modeling by the shortest data description},''
\newblock {\em Automatica}, vol. 14, pp. 465--471, 1978.

\bibitem{hurvich89}
C.~M. Hurvich and C.-L. Tsai,
\newblock ``{Regression and time series model selection in small samples},''
\newblock {\em Biometrika}, vol. 76, no. 2, pp. 297--307, 1989.

\bibitem{scharf94}
L.~L. Scharf and B.~Friedlander,
\newblock ``{Matched subspace detectors},''
\newblock {\em IEEE Trans. on Signal Processing}, vol. 42, no. 8, pp.
  2146--2157, Aug. 1994.

\bibitem{basseville93}
M.~Basseville and I.~V. Nikiforov,
\newblock {\em Detection of Abrupt Changes: Theory and Application},
\newblock Prentice Hall, Englewood Cliffs, N. J., 1993.

\bibitem{myers81}
C.~Myers and L.~Rabiner,
\newblock ``{A level building dynamic time warping algorithm for connected word
  recognition},''
\newblock {\em IEEE Trans. on Signal Processing}, vol. 29, no. 2, pp. 284--297,
  Apr. 1981.

\bibitem{svendsen87}
F.~K. Svendsen and F.~K. Soong,
\newblock ``{On the automatic segmentation of speech signals},''
\newblock in {\em Proc. 1987 Int. Conf. Acoust., Speech, and Signal Processing,
  Dallas, TX}, pp. 77--80.

\bibitem{wang2004}
X.~Wang and H.~V. Poor,
\newblock {\em Wireless Communication Systems: Advanced Techniques for Signal
  Reception},
\newblock Prentice Hall PTR, 2004.

\bibitem{golub89}
G.~H. Golub and C.~F.~Van Loan,
\newblock {\em Matrix Computations},
\newblock Johns Hopkins University Press, Baltimore, 2nd edition, 1989.

\bibitem{ferguson96}
T.~S. Ferguson,
\newblock {\em A Course in Large Sample Theory (Texts in Statistical Science)},
\newblock Chapman \& Hall/CRC, 1996.

\end{thebibliography}
\bibliographystyle{Hwang}

\newpage

\end{document}